\documentclass[trackchanges,twocolumn]{aastex701}

\usepackage{amsmath}

\begin{document}

\title{First monitoring campaign of a Main-sequence Radio Pulse emitter: the case of CU\,Vir}
\shorttitle{Monitoring of CU\,Vir}
\shortauthors{Das et al.}

\correspondingauthor{Barnali Das}
\email{barnali@ncra.tifr.res.in}

\author[0000-0001-8704-1822]{Barnali Das}
\affiliation{National Centre for Radio Astrophysics, Tata Institute of Fundamental Research, Pune University Campus, Pune-411007, India}
\email{barnali@ncra.tifr.res.in}  

\author[]{Hayley Bignall}
\affiliation{Manly Astrophysics, 15/41-42 East Esplanade, Manly, NSW 2095, Australia}
\affiliation{Visiting Scientist, CSIRO, Space and Astronomy, P.O. Box 1130, Bentley, WA 6102, Australia}
\email{Hayley.Bignall@manlyastrophysics.org}  

\author[]{Andrew Zic}
\affiliation{Australia Telescope National Facility, CSIRO, Space \& Astronomy, PO Box 76, Epping, NSW 1710, Australia}
\email{Andrew.Zic@csiro.au}

\author[0000-0002-0844-6563]{Poonam Chandra}
\affiliation{National Radio Astronomy Observatory, 520 Edgemont Road, Charlottesville VA 22903, USA}
\email{pchandra@nrao.edu}  

\author[]{Joshua Pritchard}
\affiliation{Australia Telescope National Facility, CSIRO, Space \& Astronomy, PO Box 76, Epping, NSW 1710, Australia}
\email{Joshua.Pritchard@csiro.au}

\author[]{John Morgan}
\affiliation{CSIRO, Space and Astronomy, P.O. Box 1130, Bentley, WA 6102, Australia}
\email{John.Morgan@csiro.au}

\author[]{Ankita Ghosh}
\affiliation{National Centre for Radio Astrophysics, Tata Institute of Fundamental Research, Pune University Campus, Pune-411007, India}
\email{ankita@ncra.tifr.res.in}  

\author[]{Bhaswati Bhattacharyya}
\affiliation{National Centre for Radio Astrophysics, Tata Institute of Fundamental Research, Pune University Campus, Pune-411007, India}
\email{bhaswati@ncra.tifr.res.in}

\author[]{George Hobbs}
\affiliation{Australia Telescope National Facility, CSIRO, Space \& Astronomy, PO Box 76, Epping, NSW 1710, Australia}
\email{George.Hobbs@csiro.au}

\begin{abstract}
CU\,Vir, a magnetic hot star, is the first discovered Main-sequence Radio Pulse emitter (MRP) characterized by its ability to produce periodic radio pulses via electron cyclotron maser emission. Although significant advancements have been made in understanding MRPs, their temporal properties remain mostly unexplored. 
To overcome this limitation, we conducted a pilot study with the Australia Telescope Compact Array, in which we observed pulses from CU\,Vir at 36 epochs over 1--3 GHz. In this frequency range, CU\,Vir produces two $\approx 100\%$ circularly polarized pulses, called `leading' and `trailing' pulses per rotation period. 
We find significant differences in the variability indices exhibited by the two pulses as a function of frequencies, with the leading pulse showing higher variability throughout our observing band. 
This result could be explained in the scenario of centrifugal breakout events in the magnetosphere of an oblique rotator causing correlated fluctuations across frequencies, along with intrinsic instabilities associated with coherent emission.
In addition, we discover jittering in the arrival phases of pulses that must be considered in future monitoring campaigns.
The pulses also exhibit a systematic shift to later arrival times during the course of our observing campaign, allowing us to refine the rotation period to $0.5206882$  days. 
Finally, we estimate that $\sim 30$ pulses will be needed to extract global pulse properties for the leading or trailing pulses. This relatively small number strongly motivates more extensive monitoring campaigns of MRPs, both to validate our results, and also to pinpoint the origin of the observed temporal variations.
\end{abstract}



\section{Introduction}\label{sec:intro}
Main-sequence Radio Pulse emitters (MRPs) are stellar radio variables characterised by their ability to produce periodic radio pulses by the coherent electron cyclotron maser emission (ECME) mechanism \citep{das2021}. They are known to be main-sequence stars of spectral types B and A that harbour large-scale, kG-strength surface magnetic fields \citep[e.g.][]{trigilio2000}, which can often be approximated as dipoles inclined to the stellar rotation axes \citep[e.g.][]{shultz2018}. These magnetic fields interact with their radiatively driven wind to form large-scale, co-rotating magnetospheres \citep[e.g.][]{townsend2005,petit2013,owocki2016}. For the early B-type stars, the magnetospheres manifest over a wide range of the electromagnetic spectrum spanning radio to X-ray \citep[][etc.]{petit2013,leto2018}. However, for the late B and A type magnetic stars, radio is the primary waveband where magnetospheric signatures are unambiguously observed \citep[e.g.][]{shultz2020,das2024}.


Among the different magnetospheric emission, ECME is unique in several aspects. It is a beamed emission and the beaming geometry is related to the magnetic field direction; the frequency of emission is proportional to the local magnetic field strength \citep[][etc.]{melrose1982,lim1996,treumann2006,trigilio2011}. All these properties make it the only probe capable of providing information on highly dynamic events that occur on small spatial scales \citep{das2021, polisensky2023}. The same set of properties also enable us to infer magnetic field strengths and topologies of objects, as well as 3D magnetospheric plasma distribution using ECME observations at multiple frequencies \citep{das2020a,das2024,kavanagh2024}. Thus, ECME has been demonstrated to be a versatile magnetospheric probe, but to achieve that, it is important to have a thorough understanding of the phenomenon.

With well-characterized and stable magnetic properties \citep[achieved through optical spectropolarimetry, e.g.][]{shultz2018}, MRPs are excellent targets to fully characterize ECME.
As part of that, 
it is important to quantitatively understand how ECME properties are related to stellar magnetospheric parameters, as well as how they vary with time and frequency.
The biggest challenge in answering these questions is the lack of adequate data. MRPs have rotation periods ranging from several hours to a few days, and their pulses have widths $\sim$1\,hour \citep[e.g.][]{das2022a}.
Consequently, constructing a statistically significant sample of MRP pulses is expensive in terms of telescope time. Despite that, significant progress has been made in expanding the MRP sample size, and understanding spectral evolution in recent years \citep[e.g.][]{leto2016,das2020a,das2022a,das2025a}.
However, the one question that has been 
least explored is the temporal evolution of ECME. Most MRPs have been observed only at single epochs, and so far, only one MRP has been observed at more than two epochs \citep[CU\,Vir, the first discovered MRP,][]{trigilio2000,trigilio2008,ravi2010,trigilio2011,lo2012,das2021}. However, these studies primarily focus on spectral evolution, and barring \citet{das2021}, any temporal changes in the pulse strength/profile are attributed to intrinsic instability at the emission sites, while changes in pulse arrival phases are attributed to rotation period evolution of the star (CU\,Vir). \citet{das2021}, on the other hand, reported correlated changes in the peak flux densities for pulses at different frequencies, and even for pulses produced at opposite magnetic hemispheres observed at two epochs separated by a year, which led them to speculate that the ECME phenomenon is intrinsically stable, and the underlying cause of observed temporal variation is external to the emission sites.

All these results provide strong motivation to conduct dedicated monitoring campaigns for MRPs to reveal the true scenario(s) regarding the nature and cause of temporal evolution.
As a first step towards that goal, we conducted a pilot project in which we observed pulses from CU\,Vir at 1--3 GHz at 36 epochs with the ATCA. 
In this paper, we report our initial results from this monitoring campaign.

This paper is structured as follows: in the next section (\S\ref{sec:cuvir}), we describe our target CU\,Vir. This is followed by observation and data analysis (\S\ref{sec:data}) and results (\S\ref{sec:results}). We discuss our key findings in \S\ref{sec:discussion} and summarize the paper in \S\ref{sec:summary}.

\section{Our target: CU\,Vir}\label{sec:cuvir}
ECME from CU\,Vir was first discovered by \citet{trigilio2000}. They observed two 100\% right circularly polarized (RCP) pulses at their lowest frequency of observation (1.4 GHz), but no counterparts at 5 GHz and above. They also noted that the arrival phases of the pulses nearly correspond to the zeros of the star's longitudinal magnetic field curve, suggesting that the emission is beamed at approximately a right angle to the magnetic dipole axis. Based on these properties, the emission mechanism was identified as ECME \citep{trigilio2000,trigilio2008}.

Following this discovery, a number of studies were conducted to understand the pulse properties \citep{trigilio2008,ravi2010,trigilio2011,lo2012,das2021}. These studies established that CU\,Vir produces ECME of both circular polarizations (associated with opposite magnetic hemispheres), but the two have different spectral extents. In particular, above 1 GHz, ECME is predominantly produced from one magnetic hemisphere only, leading to the observation of $\approx 100\%$ circularly polarized pulses up to around 3 GHz \citep{das2021}. 
The LCP pulses are dominant at sub-GHz frequencies ($\lesssim400$\,MHz to 800\,MHz) and exhibit cut-off at $\approx 1.4$ GHz \citep{das2021}.
The two RCP pulses observed above 1 GHz are termed leading and trailing pulses \citep[the pulses observed around rotational phases 0.3 and 0.7 respectively in the Figure 2 of][]{das2021}.

The multi-epoch studies of CU\,Vir revealed a curious phenomenon not observed for any other MRPs. After the discovery of the pulses at 1.4 GHz (20\,cm band), \citet{trigilio2008} first attempted to observe it at a different radio band (13 cm, with a bandwidth of 128 MHz). While they detected the counterpart to the trailing pulse, no such counterpart was found for the leading pulse. The 13 cm leading pulse was eventually detected by \citet{ravi2010}, but again disappeared in the subsequent observation reported by \citet{lo2012}. The reason behind the intermittent nature is believed to be intrinsic instability associated with any coherent mechanism. However, no explanation exists regarding why such instabilities affect only the leading pulse and not the trailing pulse, even though both are produced by the same magnetic hemisphere.

Another important characteristic of CU\,Vir relevant for this study is its peculiar rotation period evolution. Magnetic hot stars are expected to gradually spin-down with time \citep{ud-doula2009}. 
However, multi-waveband data of CU\,Vir suggest that
it exhibits a nearly sinusoidal evolution of its rotation period, marked by alternative spin-up and spin-down \citep{mikulasek2011,mikulasek2019}.
Unfortunately, this model of rotation period evolution has been found to be inadequate to align radio pulses from different epochs \citep{das2021}\footnote{We expect the pulses at a given frequency to arrive at the same rotational phases irrespective of the epoch of observation}. 
In this paper, we use the ephemeris of \citet{mikulasek2011}, which was used in the most recent radio study of CU\,Vir \citep{das2021}.

CU\,Vir has a rotation period of $\approx 12$ hours \citep{mikulasek2011}. In order to reduce the required telescope time for our monitoring campaign, we observed CU\,Vir only over narrow rotational phase ranges, that were predicted to contain the arrival phases of the pulses \citep[based on past radio observation, and using the ephemeris of][]{mikulasek2011}.

\section{Observation and Data analysis}\label{sec:data}
We observed CU\,Vir with the Australia Telescope Compact Array (ATCA) with its 16-cm band (1--3 GHz) at 37 epochs between March 31, 2024 and September 28, 2024. 
The data from March 31, 2024 suffered from technical issues, leaving us with 36 epochs.
We used the Compact Array Broadband Backend (CABB) that provides an instantaneous bandwidth of 2 GHz (1--3 GHz). This set-up is different from that used in past ATCA observations of CU\,Vir, involving two bands centred at 1.384 GHz and 2.638 GHz, each of bandwidth 128 MHz \citep[e.g.][]{ravi2010}.
Each slot was of length 3 hours (including overheads) intended to cover one of the two radio pulses from the star. We observed 1934--638 as the flux density and bandpass calibrator at all but four epochs. For the four epochs, we used 0823--500 because the preferred 1934--638 was below the horizon. 1351--018 was used as the phase calibrator at all epochs.


The data were calibrated using standard procedure in \textsc{miriad} \citep{sault1995}. Due to the poor uv-coverage of individual data-set (which arises both from the short integration time and the nearly zero declination of our target\footnote{In addition, at several epochs, the array was in East-West configuration, which exacerbated the uv-coverage.}), we adopted the following strategy so as to extract the lightcurves without imaging:
\begin{enumerate}
    \item From the epochs that contain 1934--638 as the flux and bandpass calibrator, we chose the epochs with the least contamination from radio frequency interference, and the ones without any technical issues during observation. That led to 28 epochs spanning April 1, 2024 to September 22, 2024.
    \item For each epoch, we identified the timeranges where the target varies significantly by examining the calibrated Stokes V visibilities.
    \item Data from these epochs, excluding the timeranges of flux density enhancements in the target, were concatenated. 
    \item This single data set was self-calibrated (phase-only) to get a constant sky model. This was done using \textsc{wsclean} \citep{offringa2014}.
    \item After self-calibration, we made 10 images by randomly choosing any 5 out of the 28 epochs for each. This was done to verify that that target flux densities are indeed constant (taking into account of 10\% absolute flux density uncertainty).
    \item The constant sky model was used to self-calibrate all epochs (phase-only) containing the full on-source time.
    \item The model was subtracted from the corrected data to get the residual visibilities.
    \item The residual visibilities were used to construct dynamic spectra in Stokes I and V by averaging over all baselines. 
\end{enumerate}

Note that since we subtracted a constant sky model, including our target source, the average basal flux densities in both Stokes I and V are expected to be zero.

\section{Results}\label{sec:results}
\begin{figure*}
    \centering
    \includegraphics[width=0.4\textwidth]{ 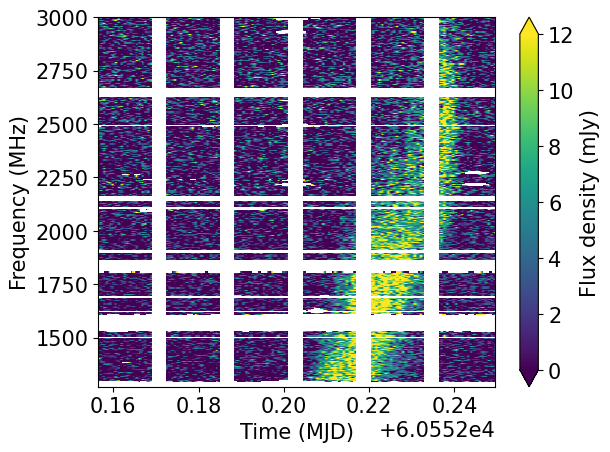}
    \includegraphics[width=0.4\textwidth]{ 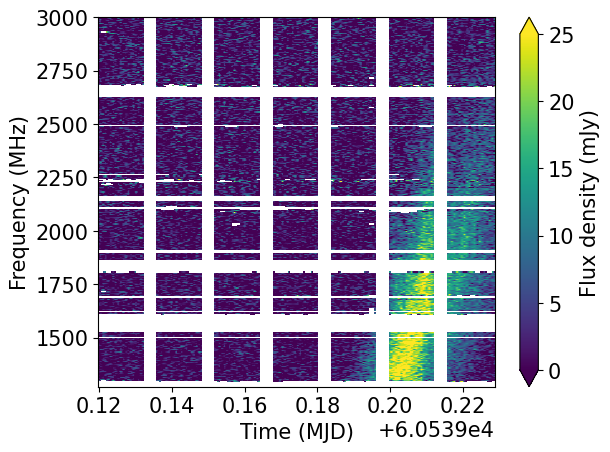}
    \includegraphics[width=0.4\textwidth]{ 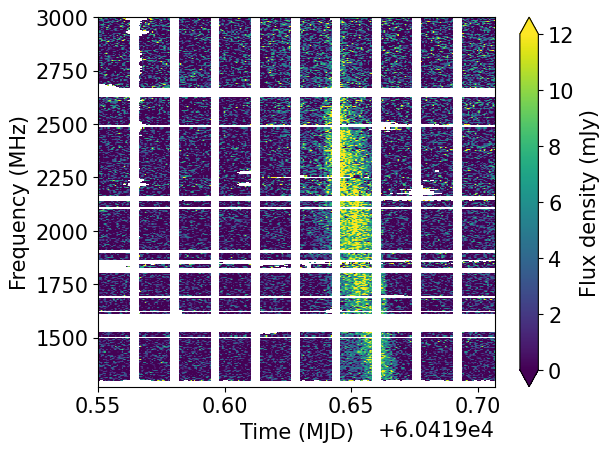}
    \includegraphics[width=0.4\textwidth]{ 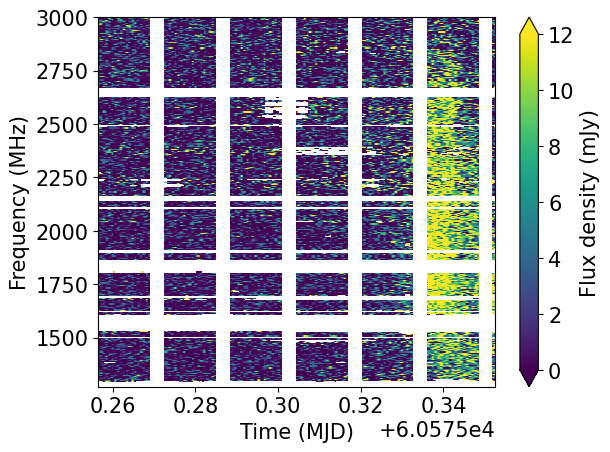}
    \caption{Dynamic spectra (Stokes V) at four epochs, the time resolution is 1 minute and the spectral resolution is 5 MHz. The epochs in the top panel correspond to the leading pulse and those in the bottom panel correspond to the trailing pulse. 
    \label{fig:example_dynamic_spectra}}
\end{figure*}

As mentioned already, each individual session was of duration only 3 hours (including overheads), and the suitable observing slots, which should contain one of the pulses, were calculated using the ephemeris of \citet{mikulasek2011}. However, we discovered that at our observing epoch, the rotational phases of arrival of the pulses are offset from those in 2021 \citep{das2021} by more than 0.05 rotation cycles ($\gtrsim 40$ minutes). This, added with the fact that some of our observations were affected by technical issues, led to the result that we were able to observe the complete pulse over the full observing band for only a subset of the observing epochs. 
This aspect is demonstrated through the four dynamic spectra (Stokes V) in Figure \ref{fig:example_dynamic_spectra}. The two epochs plotted in the top panel correspond to the observation of the leading pulse and the ones in the bottom correspond to the observation of the trailing pulse. As can be seen, for the epochs on the left, our observing windows contain the complete pulse throughout the band, but for the ones on the right, only a fraction of the band contains the full pulse. A manuscript focussed on detailed characterization of the dynamic spectra is under preparation (Fergusson et al. in prep.).

In the subsequent sections, we present the lightcurves and spectra obtained at different epochs, and compare the temporal variations observed for the leading and trailing pulses. We use Stokes V data instead of Stokes I since the pulses are known to be $\approx 100\%$ circularly polarized at these frequencies and the noise in Stokes V is lower than that in Stokes I due to lower residual source confusion. Thus, unless mentioned otherwise, all results correspond to Stokes V in the rest of the paper.

\subsection{Extraction of lightcurves}\label{subsec:lightcurves}
A number of past ATCA observations of pulses from CU\,Vir were carried out with two discrete bands centred at 1.4 GHz and 2.5/2.6 GHz with a bandwidth of 128 MHz at each band \citep{trigilio2008,ravi2010}. To enable comparison with those lightcurves, we divided our band into sub-bands of width 128 MHz and centred at frequencies 1.140 GHz, 1.268 GHz, 1.396 GHz, 1.524 GHz, 1.652 GHz, 1.780 GHz, 1.908 GHz, 2.036 GHz, 2.164 GHz, 2.292 GHz, 2.420 GHz, 2.548 GHz, 2.676 GHz, 2.804 GHz, 2.932 GHz and 3.06 GHz. 
For each sub-band, lightcurves were extracted with a time resolution of 2 minutes \citep[in accordance with][]{ravi2010}. The key results obtained by examining the lightcurves are described below.

\subsubsection{Evolution of rotational phases of arrival of the pulses}\label{subsubsec:phi_rot_evolution}
\begin{figure*}
    \centering
    \includegraphics[width=0.3\textwidth]{ 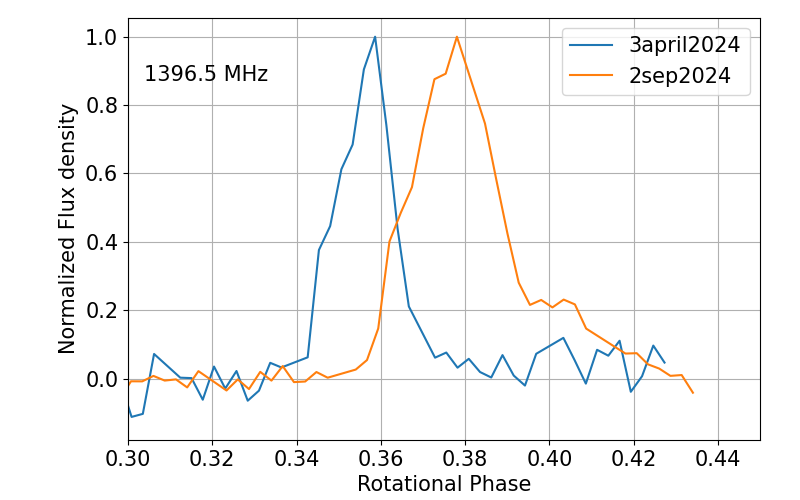}
    \includegraphics[width=0.3\textwidth]{ 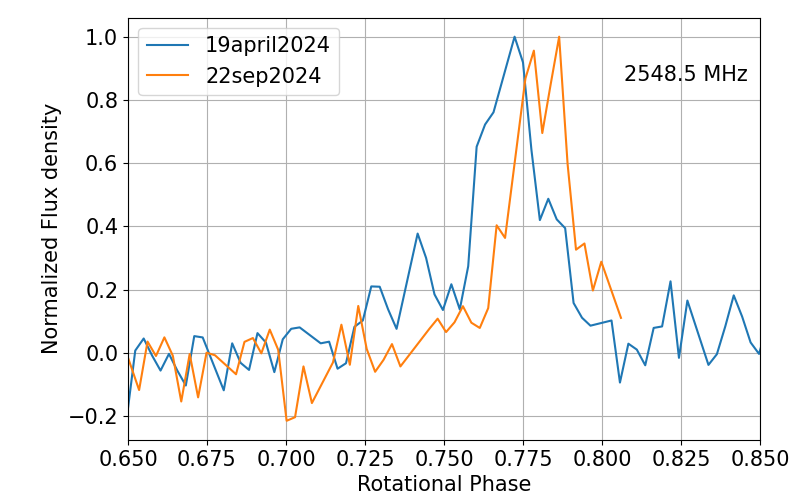}
    \includegraphics[width=0.3\textwidth]{ 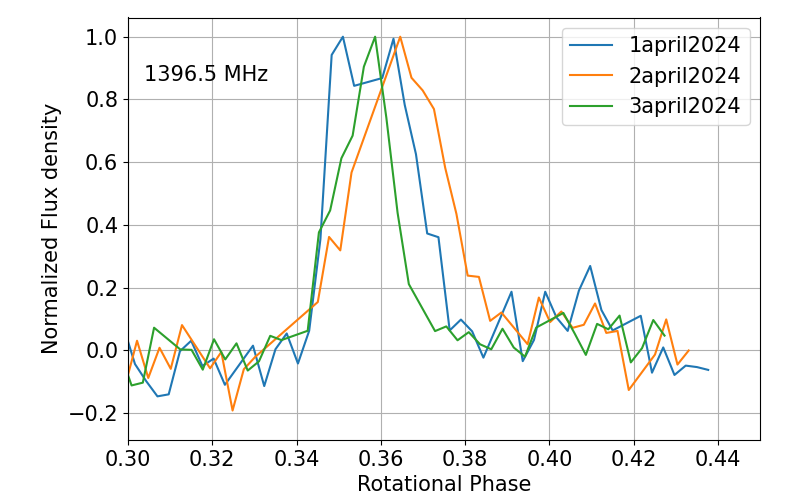}
    \caption{\textit{Left and Middle:} Comparison of the pulse arrival times for the leading (1.4 GHz) and trailing (2.5 GHz) respectively at two epochs separated by similar timescales.
    \textit{Right:} Comparison of the pulse arrival windows for the leading pulse at 1.4 GHz on three consecutive days demonstrating the pulse arrival times `jitter'.}
    \label{fig:pulse_jittering}
\end{figure*}

Figure \ref{fig:lc_leading} and Figure \ref{fig:lc_trailing} show the lightcurves at 1.4 GHz and 2.5 GHz (central frequencies of previous ATCA L and S bands) for the leading and trailing pulses respectively. 
We find a significant evolution in the pulse-profile and in pulse strength in the different epochs \citep[also noted by past studies, e.g.][]{trigilio2011}. Besides, we find that the pulse arrival phases also evolve with time. This aspect is highlighted by the grey shaded regions that approximately mark the rotational phase windows encompassing the pulses (Figures \ref{fig:lc_leading} and \ref{fig:lc_trailing}). Note that we have used a fixed rotational phase window for a given pulse type (leading or trailing) and at a given frequency. By comparing the relative positions of the peaks of the pulses within the shaded regions, we find that pulses at later epochs arrive at later phases as compared to those at the earlier epochs (e.g. compare the leading pulses observed on 2024-04-03 with that observed on 2024-09-02, Figure \ref{fig:lc_leading}). This can be seen more clearly in the left and middle panels of Figure \ref{fig:pulse_jittering}, where we plot the pulses from two epochs on top of each other (leading on the left panel and trailing on the right panel). 
For the case of the leading pulse where both pulses have relatively simple profiles (left panel of Figure \ref{fig:pulse_jittering}), the peaks are offset by 0.019 rotation cycle ($\approx 14$ minutes) for the pulses observed at two epochs separated by 152 days.
This type of shift could arise due to the use of an incorrect ephemeris. 

In addition to the above shift in the pulse arrival phases, we also find that pulses observed on consecutive days can arrive at slightly different rotational phases. An example is shown in the right panel of Figure \ref{fig:pulse_jittering}. If we consider the falling edge of the pulses, the maximum offset observed between the pulses observed at epochs separated by just two days is $\approx 0.014$ rotation cycle or 10 minutes. This shift is unlikely to be caused by a change in the rotation period, and instead resembles phase-jitter observed for pulsars \citep[e.g.][]{rathnasree1995}. Thus, similar to pulsars, MRP pulses also exhibit jitter in intensity, profile (already known) and arrival phases (this work).

\begin{figure*}
    \centering
    \includegraphics[width=0.9\textwidth]{ 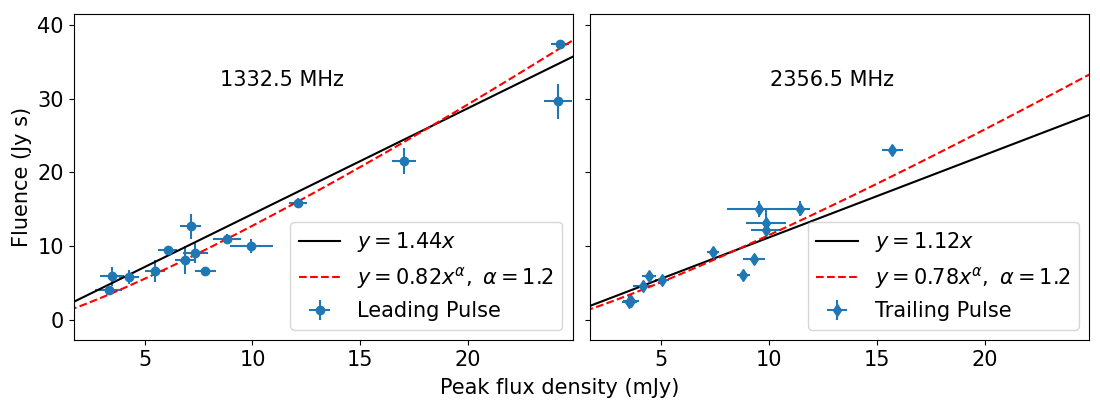}
    \caption{Variation of the fluence (Eq. \ref{eq:fluence}) with peak flux densities at different epochs for the leading (\textit{\textbf{Left}}) and the trailing (\textit{\textbf{Right}}) pulses. Note that we have converted the rotational phase bins to time-bins to use a more conventional unit for fluence.
    \label{fig:fluence}}
\end{figure*}

\subsubsection{Evolution of fluence}\label{subsubsec:fluence}
We define the fluence of the pulses as:
\begin{align}
    F&=\int S d\phi_\mathrm{rot},\label{eq:fluence}
\end{align}
where $S$ is the flux density at a phase $\phi_\mathrm{rot}$ and the integration is carried out over the rotational phase window containing the relevant pulse. As indicated by this equation, the calculation of fluence requires complete coverage of the pulse under consideration. However, overall the pulses were found to arrive at later times than that observed for the pulses reported by \citet{das2021}. This resulted into the fact that for the leading pulse, which drifts to higher frequencies with time, we do not have the full pulse coverage at several epochs at the upper half of the observing band. In case of the trailing pulse that drifts in the opposite direction on the frequency-time plane, we do not have the full pulse coverage at the lower half of the band for a number of epochs. In order to maximize the number of epochs that enable fluence calculation, we carry out this exercise at a single frequency bin of width 512 MHz (to maximize the SNR) with central frequencies of 1332.5 MHz (lowest frequency bin) for the leading and 2356.5 MHz (second highest frequency bin as the pulse is very weak at the highest frequency bin) for the trailing pulses.

Figure \ref{fig:fluence} shows the epoch to epoch variation of fluence as a function of peak flux densities at two representative frequencies for the leading and trailing pulses (15 and 13 epochs for the leading and trailing pulses respectively).
For both pulse-types, the fluence and the peak flux density are proportional to each other and are approximately related by a linear relation.  This observation indirectly suggests that the duration (width) of the pulses does not evolve with time.

It is interesting to note that the slope of the straight line fitted to the observed data points (fluence vs peak flux density) are near identical for the leading and trailing pulses ($1.44\pm0.05$ and $1.12\pm0.08$ respectively). By varying the frequency bin (excluding the highest bin), we recover the same slope for both pulse types, showing that the pulse duration does not evolve significantly with frequency or pulse type between 1 to 2.6 GHz.

This exercise establishes the fact that the peak flux density and the fluence are positively correlated so that one can be used as a proxy for the other. This allows us to use peak flux density instead of fluence when comparing pulse energies at different epochs.

\subsubsection{Average pulse-profile}\label{subsubsec:average_profile}
\begin{figure*}
    \centering
    \includegraphics[width=0.8\textwidth]{ 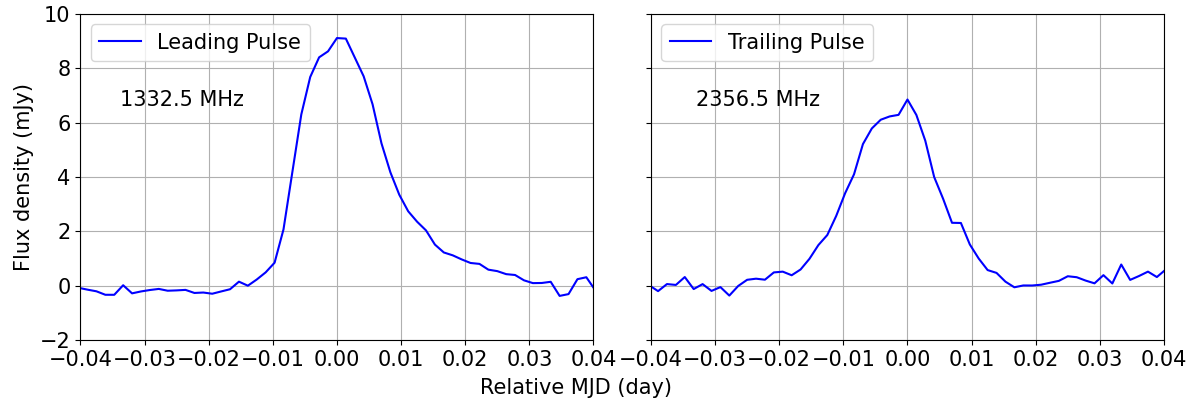}
    \caption{Pulse profiles obtained by stacking lightcurves for the different observing epochs. The \textbf{\textit{Left}} and the \textbf{\textit{Right}} panels correspond to the leading and trailing pulses respectively. For the leading pulse profile, 15 pulses were stacked and for the trailing pulse, 13 pulses were stacked. The time resolution is 2 minutes and the bandwidth is 512 MHz around the central frequencies shown in the figure. \label{fig:average_profiles}}
\end{figure*}

We finally examine the average pulse-profiles obtained from the different epochs of observations.
Here also, we ideally require the complete coverage of the pulses for the epochs that will be used for stacking. Hence, following the argument used for fluence estimations, we divided our observing band into four sub-bands (each of width 512 MHz) and chose the frequency bins with central frequencies of 1332.5 MHz and 2356.5 MHz for the leading and trailing pulses respectively. Since, we observed pulses at different epochs not to align fully in rotational phases (both due to phase-jitter and potentially for an incorrect ephemeris), we first aligned the pulses from different epochs through cross-correlation, and stacked them in relative time rather than rotational phases. The resulting average pulse-profiles are shown in Figure \ref{fig:average_profiles}. For the leading pulse, 15 epochs could be used, and for the trailing pulse, 13 epochs could be used for this exercise (for the remaining epochs, the pulses were only partially covered). 

With the small number of epochs used, these profiles are unlikely to represent a global characteristic of the MRP. The requirements to extract global pulse-profiles will be discussed in \S\ref{subsec:average_properties}.

\subsection{Extraction of peak flux density spectra}\label{subsec:spectra}

Using the lightcurves extracted for each sub-band of width 128 MHz and with a time resolution of 2 minutes (see \S\ref{subsec:lightcurves}), the peak flux density spectra are obtained in the following way:

\begin{enumerate}
    \item For each epoch, the frequency range for which the peak of the ECME pulse lies within our observing window was visually identified. Note that due to the limitation of the existing ephemeris, we missed the peak for several epochs for which our flux calibrator 1934-638 was not up at the beginning and had to be observed at the end. 
    \item The frequencies over which the peak of the pulse was not covered, were excluded from subsequent analysis.
    \item We extracted the peak flux density spectra in two ways. First, we simply obtained the peak flux density in each lightcurve, the standard deviation in the lightcurve away from the phases of enhancement was assigned as the uncertainty in the peak flux density estimate at that frequency. 
    \item The above method yields relatively noisy spectra, especially for the frequencies where the enhancement is weak. To better fit the lightcurve peaks and reduce the noise in the spectra, each lightcurve was smoothed by sliding a convolving window of size 5. The peak flux density was taken to be the maximum observed flux density in this smoothed lightcurve. The corresponding uncertainty is obtained by scaling the standard deviation of the lightcurve away from the phases of enhancement by a factor of $1/\sqrt{5}$ (where 5 is the width of the smoothing window).
\end{enumerate}

\begin{figure}
    \centering
    \includegraphics[width=0.45\textwidth]{ 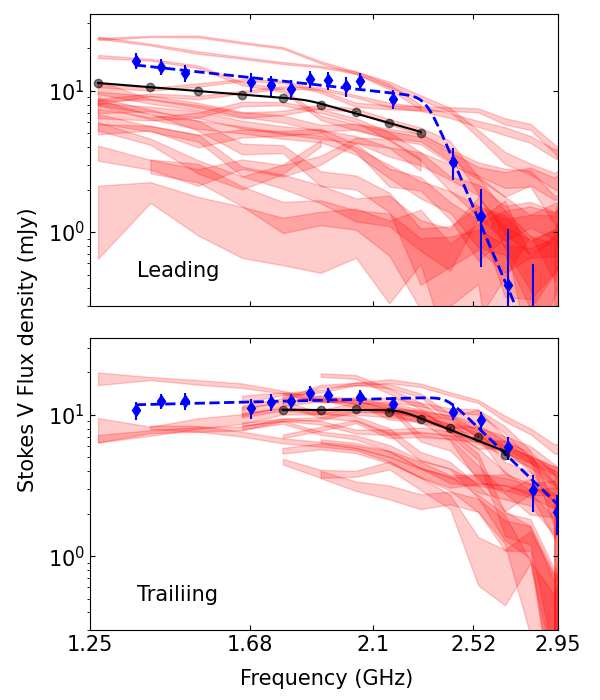}
    \caption{Comparison among the peak flux density spectra (Stokes V) observed at different epochs. The top and bottom panels correspond to the leading and trailing pulses respectively. The red shaded regions mark the spectra at individual epochs (the width of the shaded regions corresponds to $1\sigma$ uncertainty). The black markers and the solid line show the values for the time-averaged spectra and the corresponding fitted spectra. Note that the averaging is done only for the frequencies for which measurements are available at least 10 epochs. For comparison, we have also plotted the Stokes V spectra observed by \citet{das2021} on 2019--06--12 with the VLA (shown by diamonds in blue). The fitted spectra are also shown in blue dashed lines.
    \label{fig:compare_peak_flux_spectra_stokesV}}
\end{figure}

The peak flux density spectra for the individual epochs obtained using both strategies are shown in Figures \ref{fig:spec_leading} and \ref{fig:spec_trailing}.
The primary difference between the spectra obtained with the `raw' flux densities vs the ones obtained after smoothing the lightcurves is that the former have higher error bars (as expected).
In Figure \ref{fig:compare_peak_flux_spectra_stokesV}, the spectra for the leading and trailing pulses (red shaded regions) are compared among the different epochs. For clarity, here we used the spectra obtained after smoothing the lightcurves. By comparing the top and bottom panels, it appears that the leading pulse exhibits much higher levels of fluctuations in flux densities than that for the trailing pulse. In addition, we also show the Stokes V spectra (blue diamonds) reported by \citet{das2021} using the Karl G. Jansky Very Large Array (VLA). Note that for the VLA data, both leading and trailing pulses were observed within the same observing session (on 2019--06--12). The flux densities observed with the VLA lie within the flux density ranges observed with the ATCA.

We next construct the average spectra for the leading and trailing pulses using the ATCA data. 
The primary motivation for constructing the time-averaged spectra, despite the fact that we do not have a large enough sample, is to explore whether there are characteristic spectral shapes for the leading and trailing pulses of CU\,Vir. The minimum number of epochs required to extract stable properties of the star is discussed in \S\ref{subsec:average_properties}. Assuming that characteristic spectral shapes exist, the flux densities across frequencies for a given epoch are expected to be correlated. For such a scenario, it is required that the same set of epochs are used across the frequency range under consideration. Imposing this constraint, we were able to obtain an average spectrum over 1.3--2.3 GHz for the leading pulse and 1.8--2.7 GHz for the trailing pulse using 12 epochs in each case.
The results are shown in black circles in Figure \ref{fig:compare_peak_flux_spectra_stokesV}.

In order to quantitatively compare the average ATCA spectra with those obtained with the VLA, we fit the spectra with the following function (smoothly connected broken power-law):
\begin{align}
    S&=A\left(\frac{\nu}{\nu_\mathrm{b}}\right)^{\alpha_1}\left\{1+\left(\frac{\nu}{\nu_\mathrm{b}}\right)^\frac{1}{s}\right\}^{(\alpha_2-\alpha_1)s}\label{eq:fit_spec},
\end{align}
where $S$ is the flux density at a frequency $\nu$. The fitted parameters are $A,\,\nu_\mathrm{b},\,\alpha_1$ and $\alpha_2$. The `smoothness parameter' $s$ is kept fixed at 0.01. The best-fit values of the parameters as well as the reduced $\chi^2$ are listed in Table \ref{tab:spectral_fits}. Both ATCA and VLA data suggest that the break frequency $\nu_\mathrm{b}$ is higher for the trailing pulse than that for the leading pulse, however, the break frequencies are higher for the VLA spectra than those for the corresponding ATCA spectra. This could be an artifact of the limited spectral extent for the ATCA data. Among the four parameters that are fitted, the one of particular interest is the spectral index $\alpha_1$ below the break frequency, which defines the spectral shape away from the cut-off frequencies. The values of $\alpha_1$ obtained from the ATCA and VLA data (for a given type of pulse) are consistent within uncertainties, supporting the idea that the flux densities across frequencies are correlated. This idea has important consequences and should be checked in the future using a larger sample of spectra.

\begin{deluxetable}{cccccc}
\tablecaption{The results of fitting the average spectra obtained with the ATCA and those obtained with the VLA with the function given by Eq. \ref{eq:fit_spec}.\label{tab:spectral_fits}}
\tablehead{
\colhead{Spectrum type} &
\colhead{$A$} & 
\colhead{$\nu_\mathrm{b}$ (GHz)} &
\colhead{$\alpha_1$} &
\colhead{$\alpha_2$} &
\colhead{$\chi^2_\mathrm{red}$}
}
\startdata
ATCA leading & $8.6\pm0.2$ & $1.86\pm0.02$ & $-0.72\pm0.06$ & $-2.5\pm0.2$ & 3.0\\ 
ATCA trailing & $10.8\pm0.3$ & $2.21\pm0.04$ & $0.0\pm 0.2$ & $-3.5\pm 0.4$ & 6.5\\
VLA leading & $9.1\pm0.6$ & $2.32\pm 0.05$ & $-0.97\pm 0.18$ & $-20\pm 6$ & 0.39 \\
VLA trailing & $13.3\pm0.9$ & $2.40\pm 0.04$ & $0.2\pm0.2$ & $-8\pm 1$ & 0.55\\
\enddata
\end{deluxetable}


To quantify the epoch to epoch variation of the peak flux densities, we calculate the debiased variability index ($V_\mathrm{rms}^\mathrm{debiased}$) at the individual frequencies, which is defined as \citep[e.g.][]{sadler2006}:
\begin{align}
    V_{\mathrm{rms}}^{\mathrm{debiased}}&=\frac{1}{\langle S \rangle}\sqrt{\sigma_\mathrm{signal}^2-\sigma_\mathrm{noise}^2},\label{eq:mod_index}\\
    \sigma_\mathrm{signal}^2&=\frac{{\sum (S_i-\langle S\rangle)}^2}{N}\nonumber\\
    \sigma_\mathrm{noise}^2&=\frac{\sum \sigma_i^2}{N}\nonumber
\end{align}
where, $S_i$ is the flux density measurement with uncertainty $\sigma_i$ at the $i$th epoch, $\langle S\rangle$ is mean of the flux density measurements at the different epochs and $N$ is the total number of flux density measurements for the frequency under consideration. 
Here also, we use a common set of epochs across frequencies, which were the ones that were used to construct the time-averaged spectra.
The error bar in the measured $V_\mathrm{rms}^\mathrm{debiased}$ is estimated from the uncertainties in the peak flux densities using a Monte Carlo approach. 

\begin{figure}
    \centering
    \includegraphics[width=0.45\textwidth]{ 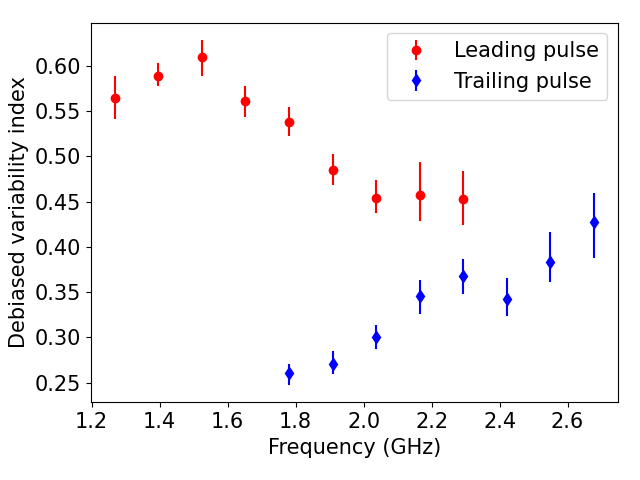}
    \caption{Epoch to epoch variation of the peak flux density spectra. The debiased variability indices are calculated using Eq. \ref{eq:mod_index} (\S\ref{subsec:spectra}).}
    \label{fig:sigma_spectra}
\end{figure}

The resulting spectral variations of the variability index for the leading and trailing pulses are shown in Figure \ref{fig:sigma_spectra}. This figure further supports the idea that the leading pulse exhibits a significantly stronger fluctuation than that exhibited by the trailing pulse. The implications of the observed trends will be discussed in \S\ref{subsec:variability}.

We do not find any `outlier' spectra in the sample either for the leading or the trailing pulses including the epochs where the pulses at 2.5 GHz were not detected. Thus, the past observations of the disappearance of the leading pulse at 13 cm is actually a consequence of a broadband phenomenon where the flux densities are suppressed across a wide frequency range.

\subsection{Significance of the observed variation of fluctuations with frequency and pulse-type}\label{subsec:variability}
\begin{figure*}
    \centering
    \includegraphics[width=0.8\textwidth]{ 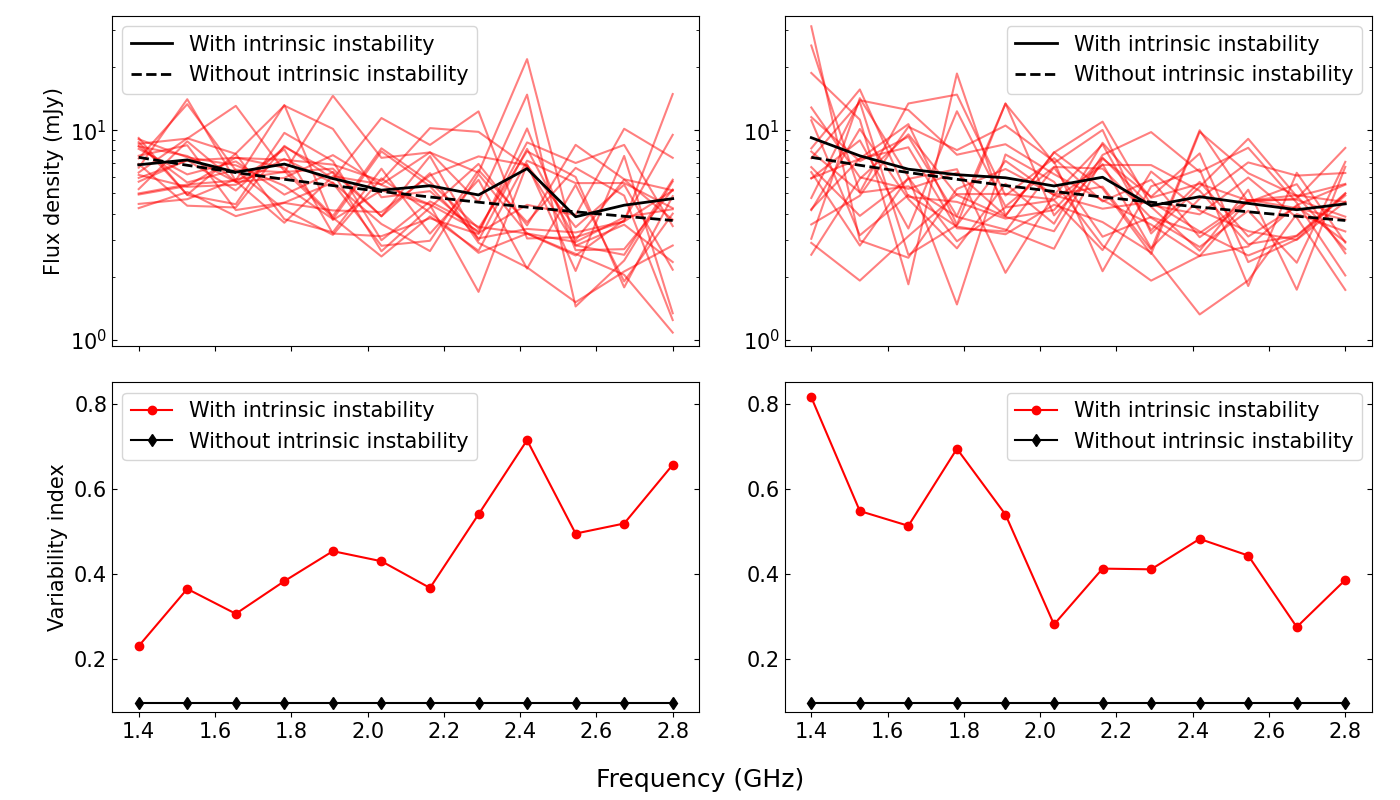}
    \caption{Simulated spectra and the corresponding variability indices. \textbf{Left} panels correspond to a scenario in which the intrinsic instability increases with increasing frequency and the \textbf{right} panels correspond to a scenario in which the instability decreases with increasing frequency. \textbf{Top} panels show the simulated spectra in red (given by Eq. \ref{eq:with_intrinsic_instability}), the black solid line shows the average of these spectra, and the black dashed lines show the average spectrum when there is no intrinsic instability (Eq. \ref{eq:no_intrinsic_instability}). \textbf{Bottom} panels show the corresponding variability indices, red circles mark the variability indices obtained from the spectra that are affected by intrinsic instability, and the black diamonds show the corresponding values for the case when there is no intrinsic instability (Eq. \ref{eq:var_index_constant}).
    \label{fig:sim_spec_var_ind}}
\end{figure*}

Both analyses using lightcurves and spectra of CU Vir suggest that there is a statistical difference between the degree of variability in the peak flux density (and hence fluence) exhibited by the leading and the trailing pulses, with the former showing significantly higher fluctuations (Figure \ref{fig:sigma_spectra}). The difference is highest at the lowest frequency bin and vice-versa. 


For the trailing pulse, there is a clear increase in the variability index (from $\approx 0.25$ to $\approx0.40$) between 1.8--2.7 GHz (Figure \ref{fig:sigma_spectra}). 
On the other hand, for the leading pulse, the variability index decreases from $\approx0.60$ to $\approx 0.45$ between 1.5--2.3 GHz (Figure \ref{fig:sigma_spectra}). Between 1.3--1.5 GHz, the variability index remains nearly constant (within uncertainties) with frequency. We first discuss whether the idea of the existence of a characteristic spectral shape could be consistent with the observed trend of variability indices. This is examined via a simplistic simulation, described below. Note that we do not attempt to reproduce the observed spectral properties quantitatively. We will also ignore measurement uncertainties and set $\sigma_\mathrm{noise}=0$ while calculating the variability indices.

Let us assume that the characteristic spectral shape is a power-law with a spectral index $\alpha$. The flux densities exhibit temporal evolution, which is \textit{independent} of frequency. 
We choose the following functional form to describe this component of temporal variation:
\begin{align*}
    S(t)&=S_0+S_1\sin(\tilde{t})
\end{align*}
Where $S(t)$, $S_0$ and $S_1$ have units of flux densities, and $\tilde{t}$ is dimensionless time representing different observing epochs. Thus, the functional form of the spectrum at a frequency $\nu$ and observing epoch $\tilde{t}$ is given by:
\begin{align}
    S(\nu,t)&=\nu^\alpha\{S_0+S_1\sin(\tilde{t})\} \label{eq:no_intrinsic_instability}\\
            &=f(\nu)g(t)\nonumber
\end{align}
For the above set of spectra, the variability index at any frequency $\nu$ is given by:
\begin{align}
    V^0_\mathrm{rms}&=\frac{1}{\langle g\rangle}\sqrt{\frac{{\sum (g_i-\langle g\rangle)}^2}{N}}, \label{eq:var_index_constant}
\end{align}
$N$ is the number of epochs. As expected, $V^0_\mathrm{rms}$ is independent of frequency. 

We now consider the case where there is also intrinsic instability for which the time and frequency dependences are entangled\footnote{Although the motivation for introducing this component is intrinsic instability, it could also account for any types of variation (other than instrumental noise) for which the time and frequency parts cannot be separated.}. 
We choose the functional form $\exp\{-\delta(\nu,t)\}$ to describe this component, where $\delta$ has a Gaussian distribution with a mean of $0$ and standard deviation $\sigma(\nu)$; $\delta$ is drawn randomly from this distribution for a given epoch. Thus, the final functional form of the simulated spectra are given by:
\begin{align}
    S(\nu,t)&=\nu^\alpha\{S_0+S_1\sin(\tilde{t})\}\exp(-\delta) \label{eq:with_intrinsic_instability}
\end{align}

We set $\alpha=-1$, $S_0=10$ mJy, $S_1=1.5$ mJy. The frequency is varied between 1.4 and 2.8 GHz and the number of epochs is set to 20, the epoch indices (values of $\tilde{t}$) are drawn randomly between 0 and 200 (considering that our observations are randomly distributed over a span of 181 days). For $\sigma$, we consider two scenarios: in one, we set $\sigma=0.3(\nu/\nu_0)$ and in the other, we set $\sigma=0.7(\nu_0/\nu)$, $\nu_0=1.4$ GHz. Thus, in the former $\sigma$ increases with frequency and in the other, it decreases with frequency.

Figure \ref{fig:sim_spec_var_ind} shows the results of the simulation. The left (right) panels correspond to the scenarios where $\sigma(\nu)$ increases (decreases) with frequency; top panels show the simulated spectra and the bottom panels show the variability indices. In the top panels, the red solid lines are the simulated spectra at different epochs and the thick black solid lines show the corresponding time-averaged spectra. In addition, we also show the time-averaged spectrum in the absence of intrinsic instability in black dashed line, which is described by a power-law with spectral index of $\alpha=-1$. 

In the bottom panels, the variability indices in presence of intrinsic instability are shown in red circles, the black diamonds mark $V^0_\mathrm{rms}$, which is independent of frequency. For the scenario where $\sigma$ increases with frequency, the variability index approximately increases with frequency, and vice-versa.

It is interesting to note that for this simulation, even though the variability indices vary significantly with frequency, the time-averaged spectra (black solid lines in top panels) still closely follow the corresponding time-averaged spectra in the absence of intrinsic instability (black dashed line). This suggests that it is possible to retrieve the characteristic spectral shapes from a relatively small sample size (in this case, 20 epochs of observation) even though the variability indices are high, and vary with frequency.
Thus, the time-averaged spectra shown in Figure \ref{fig:compare_peak_flux_spectra_stokesV} do not lose their significance in view of the observed variability indices (Figure \ref{fig:sigma_spectra}), and only by further expanding the sample spectra, can one confirm whether or not they are characteristics of CU\,Vir.

Figure \ref{fig:sim_spec_var_ind} also provides insight into the implications of the observed difference between the variabilities of the leading and trailing pulses. For the trailing pulse, the observed variation suggests an increasingly unstable environment for ECME production as we go closer to the magnetic poles (where the magnetic field strength and hence the ECME frequencies are higher). 
This scenario is related to the scenarios proposed for the observation of fine structures (in seconds timescale) in ECME pulses of the MRP HD\,142990 close to the cut-off frequencies \citep{das2025b}.
The authors proposed that the seemingly smooth ECME pulse profile is created by the superposition of many spiky features (fine structures) each of which corresponds to elementary sites of ECME production in the stellar magnetosphere. With increasing frequencies, the density of these sites decreases, leading to fragmentation of the smooth envelope into the observed fine structures (eventually causing cut-off). A smaller number of emission sites (equivalently, smaller spatial scales of the emission sites) are more likely to retain signatures of fluctuations in local conditions, leading to an increased variability index at these frequencies. Note that, we do not detect any fine structures for the case of CU\,Vir which could be due to inadequate sensitivity.

For the leading pulse, which exhibits a higher level of fluctuations than that for the trailing pulse, and also an opposite trend in the relation between variability index and frequency, it is possible that this pulse is subject to additional sources of temporal variation. The above simulation suggests that this additional source introduces an instability, of which degree decreases with frequency. Although this aspect makes interstellar scintillation a plausible candidate, we rule it out considering that such effects should not distinguish between the leading and trailing pulses.
Thus, for the additional source to be able to affect only the leading pulse, its motion must be tied to that of the star's magnetosphere. The simplest explanation is propagation effects within the stellar magnetosphere. CU\,Vir is known to have a highly oblique magnetic field with respect to its rotation axis \citep{kochukhov2014}, which gives rise to an azimuthally asymmetric distribution of plasma in the magnetosphere. In addition, the magnetic field is also known to deviate from the assumed axi-symmetric geometry \citep{kochukhov2014}, which will further contribute to an asymmetric magnetosphere. These aspects can also alter the sites of CBOs and/or their distribution around the magnetic equatorial region. Both these aspects together could result in the observed difference in the temporal variations of the leading and trailing pulses.

To summarize, it is possible that ECME from CU\,Vir has characteristic spectral shapes, which can be retrieved through multi-epoch observations, despite the presence of intrinsic instabilities that affect different frequencies to different extents. These spectral shapes (which can be different for pulses observed at different rotational phases) are expected to be stable properties, and should be used
when comparing two MRPs. In addition, our simulation demonstrates how the frequency dependence of variability index could give us insights of nature of instabilities affecting the pulses. For the observed variability indices, CBOs remain key candidate to explain their spectral dependence and also the difference observed between leading and trailing pulses.

In the future, monitoring campaigns should be conducted for MRPs with a wide range of obliquities (angle between magnetic dipole and rotation axes and varying degrees of complexity in the magnetic field topologies. This will be able to confirm whether the observed difference between pulses of same circular polarization but at different rotational phases is specific to MRPs with certain set of properties or specific to CU\,Vir.

\section{Discussion}\label{sec:discussion}
The importance of monitoring radio pulses is well-recognized in the field of pulsars and is routinely done for several pulsars. Despite being the closest analogue to pulsars on the stellar main-sequence, similar investigation has not been carried out for MRPs until now. The primary challenge here is the much longer timescale involved ($\sim$ several hours) as the rotation periods are $\sim$ days. Here we report a pilot study with the first discovered MRP CU\,Vir, which is also one of the fastest rotating MRPs.

As a coherent phenomenon, pulses produced by ECME is expected to exhibit epoch to epoch variation since the growth rates are extremely sensitive to local conditions at the emission sites \citep[e.g.][]{trigilio2011}. Although the emission is produced in auroral rings that can have radii larger than that of the star, the effective emission sites that correspond to the observable pulses are compact regions, so that the temporal variations of the local conditions (such as the properties of non-thermal electron population) do not necessarily average out. In addition to changing the intensity, these changes might also perturb the emission beaming geometry causing phase-jitter. Most recently, \citet{morgan2026} pointed out that the compactness of the emission sites also make interstellar scintillation a relevant phenomenon to consider in this context. Thus, at our observing frequencies, weak scintillation could also contribute to the temporal variability.

In addition to the above types of temporal variability, \citet{das2021} suggested that centrifugal breakout (CBO), the phenomenon responsible for particle acceleration in stars like CU\,Vir \citep{shultz2022,owocki2022}, can also introduce variability in ECME. CBOs are small scale explosions present at all times around the magnetic equator of the stellar magnetospheres that trigger magnetic reconnection, and the energy released drives the production of non-thermal radio emission. For other magnetospheric phenomena (e.g. incoherent radio emission), the emission sites are large enough so that any variability due to CBOs washes out, and stable emission is observed. This is, however, not the case for ECME, where the effective emission sites are small \citep[see][]{morgan2026}, so that the dynamic CBO events can leave their imprints in the observed emission. In particular, \citet{das2021} observed a `giant pulse' at 700 MHz, during which the peak flux density was higher by an order of magnitude than the typical flux density reported from the star at any frequency or polarization. Interestingly, both RCP and LCP pulses (believed to be produced in opposite stellar magnetic hemispheres) were enhanced keeping the circular polarization fraction approximately constant to that observed at other epochs. Since the magnetic equator (the sites of CBOs) is common to both magnetic hemisphere, \citet{das2021} proposed that the giant pulse is a consequence of an unusually strong CBO explosion in the magnetosphere.

For variability introduced by CBOs, we would also expect correlation among temporal variations of flux densities observed at different frequencies at a given polarization. As mentioned already, for our frequency range of observation, ECME is primarily right circularly polarized so that Stokes V emission effectively traces emission at RCP. This predicts the existence of a characteristic spectral shapes of the pulses. In the absence of any other type of temporal variability, we will then expect the variability index to be independent of frequency. This was clearly found not to be the case (Figure \ref{fig:sigma_spectra}), the significance of which is discussed in \S\ref{subsec:variability}.

Our observations at 36 epochs reveal a surprising difference between the temporal variations exhibited by the leading, and the trailing pulses \citep[both pulses are produced at the same magnetic hemisphere, e.g.][]{trigilio2011}. 
While earlier observations already hinted at such a difference \citep{trigilio2008,ravi2010,lo2012}, we are able to statistically establish the existence of this difference and quantify it for the first time.
In addition, we also discovered that similar to pulsars, MRP pulses also exhibit `phase-jitter', which will need to be incorporated in predicting the time of arrival of the pulses. 

We do not detect any giant pulse in our campaign, which could either indicate that such events are extremely rare, or, that those are predominant only at sub-GHz frequencies. It is, however, worth noting that at sub-GHz frequencies (400--800 MHz), the star was observed at only two epochs \citep{das2021}. Thus, it is not clear whether the giant pulse was truly an outlier, or, whether the flux density ranges exhibited by the corresponding pulses are unusually wider. These possibilities can be checked by extending monitoring campaigns to sub-GHz frequencies.

In the subsequent subsections, we discuss the key results that emerged from this study.

\subsection{The curious case of the intermittent pulse at 13 cm}\label{subsubsec:leading_trailing_intermittent_nature}
As mentioned in \S\ref{sec:cuvir}, the leading pulse at 2.5 GHz (13 cm) has been known to completely disappear at certain epochs and thus to be intermittent in nature \citep[][etc.]{trigilio2008,ravi2010,lo2012}. The corresponding trailing pulse, on the other hand, has been thought to be stable in this aspect. 
However, it should be noted that all these previous studies were narrow-band observations involving two discontinuous frequency bands centred approximately at 1.4 GHz and 2.5 GHz. The lightcurves shown in Figures \ref{fig:lc_leading} and \ref{fig:lc_trailing} that were extracted at two discrete frequencies with similar frequency and time averaging, indeed give the impression that the leading pulse at 2.5 GHz is intermittent (not detected in 10 out of the 13 epochs with $5\sigma$ as detection threshold). Interestingly, for the trailing pulse as well, the pulse was not detected in 3 out of 19 epochs at 2.5 GHz (though, it was always detected at 1.4 GHz).  

The continuous frequency coverage between 1--3 GHz shed new light on this phenomenon. From the spectral analysis, we find that 2.5 GHz lies beyond the break frequency for both leading and trailing pulses. Thus, the pulses at 2.5 GHz are naturally expected to be weaker than those at 1.4 GHz. In addition, our data suggest that the leading pulse has a smaller break frequency than that for the trailing pulse. Thus, for comparable pulse-strength at 1.4 GHz, the leading pulse at 2.5 GHz will be even weaker than the corresponding trailing pulse, making the former more vulnerable to telescope sensitivity. This aspect, combined with the higher degree of variability for the leading pulse, is likely responsible for its apparent intermittent behaviour at 2.5 GHz.

Thus, we conclude that the previously reported intermittent nature of the leading pulse at 13 cm (2.5 GHz) is actually a consequence of a broadband phenomenon.

\subsection{Global pulse properties}\label{subsec:average_properties}
One of the important motivations for conducting monitoring campaigns is to extract information on global pulse properties. 
These global properties will be useful in deriving empirical relation between ECME properties and stellar magnetospheric parameters, and also to compare ECME with other magnetospheric emission. In the absence of such quantities, single epoch values are used in such studies, however, the resulting relations are found to have large scatters \citep[e.g.][]{das2025a}.

In the case of pulsars, one of the important characteristics is their global pulse-profiles, which are obtained by averaging over several tens of thousands of pulses \citep{rathnasree1995}. This profile is used in pulsar timing and also to extract information about their magnetospheres and emission mechanism \citep[e.g.][]{xu2021}.

\begin{figure*}
    \centering
    \includegraphics[width=0.45\textwidth]{ 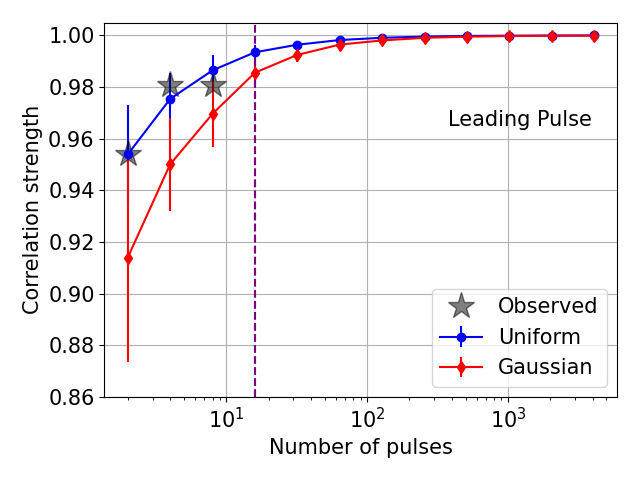}
    \includegraphics[width=0.45\textwidth]{ 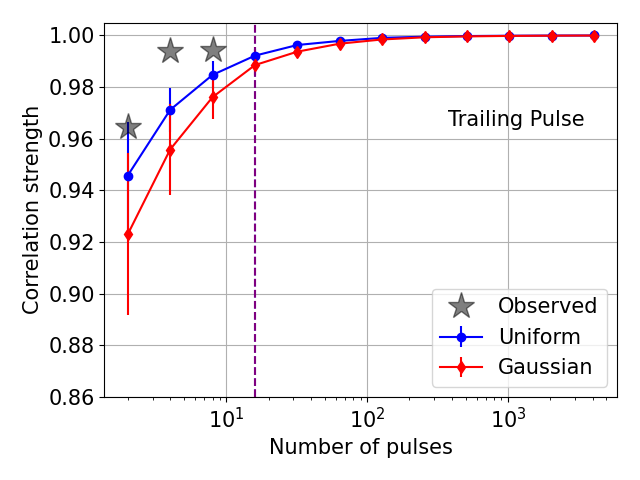}
    \caption{Correlation strength vs the number of pulses averaged for the leading (\textbf{\textit{Left}}) and the trailing (\textbf{\textit{Right}}) pulses. For details, see \S\ref{subsec:average_properties}. The dashed vertical lines mark the knees of the curves (came out to be identical for either distribution).
    \label{fig:profile_stability}}
\end{figure*}

In the case of MRPs also, the global pulse-profile, its evolution with frequencies and dependence on stellar magnetospheric parameters could give us new insights on ECME and how they are affected by the hosts.
Unfortunately, due to the much longer timescale involved, acquiring observations for thousands of pulses is not feasible. 
An interesting question in this context is the minimum number of pulses that will be needed to extract the characteristic pulse-profile. To get an estimate of this number, we employ the following strategy \citep[inspired from][]{ghosh2025}:

\begin{enumerate}
    \item We simulated $2^n$ pulses from the aligned set of observed pulses for two possible distributions of flux densities: (1) uniform distribution with minimum and maximum values corresponding to the observed minimum and maximum values of flux densities, and (2) Gaussian distribution with mean and standard deviation equal to that obtained for the corresponding observed flux densities at a given phase. We set $n=12$ (corresponding to 4096 simulated pulses for each pulse type). The flux densities are drawn randomly from these distributions.
    \item We averaged the pulses in steps of 2, thus in the $i^\mathrm{th}$ step, we averaged over $2^i$ pulses, where $i$ goes from $1$ to $12$.
    \item At each step, we calculate the normalized cross-correlation between the pulse obtained by averaging over $2^{i-1}$ pulses and that obtained by averaging over $2^i$ pulses. The maximum value of the resulting cross-correlation will be called correlation strength.
    \item The above steps were repeated 100 times to obtain the mean and uncertainties in the estimated correlation strengths.
\end{enumerate}
The correlation strengths obtained this way are plotted against the maximum number of pulses used in averaging ($2^i$ at the $i^\mathrm{th}$ step) in Figure \ref{fig:profile_stability}. Also shown are the correlation strengths for the observed set of pulses (black stars). For both leading and trailing pulses, the convergence rate to a stable profile is higher when the flux densities (at a given rotational phase) follow a uniform distribution than that for the Gaussian distribution. This is not surprising given that for the latter, the flux densities are effectively allowed to vary over a much wider range of values. This difference is especially prominent for the case of the leading pulse. Despite this aspect, all four curves are found to have their `knees' at a value of 16. Defining the number just above the knee as the minimum number of pulses required to construct a characteristic pulse profile, we find that for either leading or trailing pulses, we need at least 32 pulses to construct a global template. Although the current number of pulses available from this monitoring campaign is smaller than this value (15 and 13 for the leading and trailing pulses respectively), this number is surprisingly smaller than that used in the pulsar community suggesting that experiments aiming to construct global pulse template of MRPs are certainly feasible (will require $<100$ hours of telescope time if a strategy similar to that employed here is used for observation).

Central to this whole exercise lies the assumption that a stable profile exists. Mathematically, convergence to a stable profile is conditional, determined by the distribution of flux densities observed at different times (at a given frequency). Nevertheless, the existence of global pulse profiles for pulsars, and the general idea that the magnetospheres of MRPs are largely stable, strongly motivates to explore whether similar characteristic could exist for MRPs. 
    
A few caveats about our estimation of minimum number of pulses required for a global profile are the following:
\begin{enumerate}
     \item We made the assumption that the observed set of pulses reasonably span the flux density range that the star can emit at a given time. This may lead to an underestimation of the actual number.
    \item We treated the flux density at two consecutive phase-bins (or relative time-bin) to be uncorrelated. This need not be true and in that case, our method could lead to an overestimation of the number.
    \item Given that we observed a frequency dependence of the fluctuation (Figure \ref{fig:sigma_spectra}), the number required to construct a stable profile will likely be frequency dependent as well.
\end{enumerate}

In summary, through this first MRP monitoring campaign, we find that at least $\sim30$ pulses are needed to construct a stable profile for either leading or trailing pulses. However, this number should be treated with caution and is primarily intended to guide future monitoring campaigns, which will be critical to verify these results and also to extend it to other MRPs and variables with similar timescales such as brown dwarfs emitting periodic radio pulses \citep[e.g.][]{hallinan2007}, and some Long Period Transients \citep[e.g.][]{lee2025}.

\subsection{Implication for the rotation period model}\label{subsec:ephemeris_model}
\begin{figure}
    \centering
    \includegraphics[width=0.45\textwidth]{ 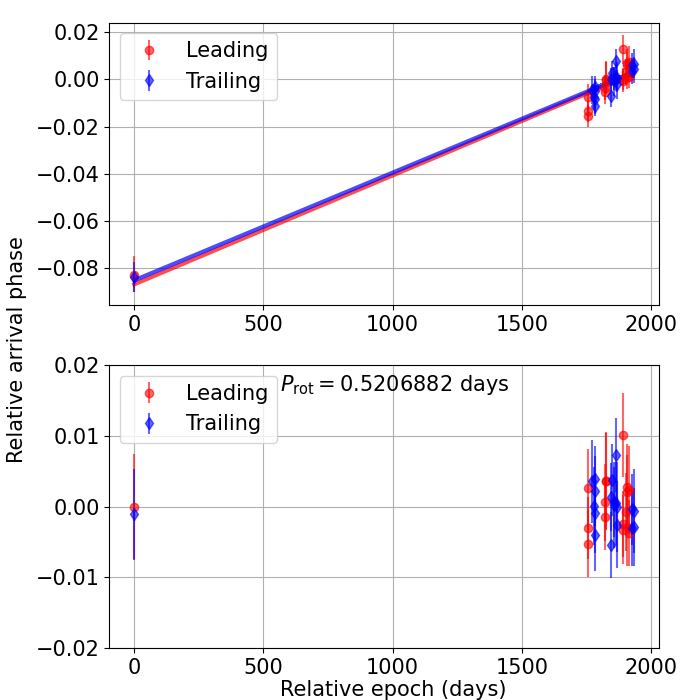}
    \caption{\textbf{Top:} Relative arrival phases calculated using the variable period ephemeris of \citet{mikulasek2011} for the pulses observed in our ATCA campaign in 2024, along with that observed by \citet{das2021} on 2019--06--12 (leftmost data points) with the VLA. 
    For the leading pulse (red circles), the arrival phases were calculated at a frequency of 1.3 GHz (ATCA data) and 1.5 GHz (VLA data). For the trailing pulse (blue diamonds), the frequencies used are 2.4 GHz (ATCA data) and 2.6 GHz (VLA data). See \ref{subsec:ephemeris_model} for details.
    The linear trends observed for both leading and trailing pulses shown by the solid lines suggest the use of an incorrect rotation period. 
    \textbf{Bottom:} Arrival phases calculated using a constant rotation period of $P_\mathrm{rot}=0.5206882$ days, which is longer than that indicated by the rotation period model of \citet{mikulasek2011}.}
    \label{fig:arrival_phase_vs_epoch}
\end{figure}
In addition to obtaining global pulse properties, another key purpose of a monitoring campaign of pulse emitting objects is to extract information on rotation period evolution. Pulsar timing is the best example of that. Using the ephemeris of \citet{mikulasek2011}, we find that pulses at later epochs arrive relatively late compared to those at earlier epochs (\S\ref{subsec:lightcurves}). Assuming that the pulse emitting regions are stable, such a case can occur if the rotation period ($P_\mathrm{rot}$) for the latter epochs is actually longer than that used while calculating the rotational phases. Note that for the variable period model of \citet{mikulasek2011}, $P_\mathrm{rot}$ decreases by less than 0.5 second during our observing session, effectively making it a constant quantity for the pulses collected during this period. The mean value of $P_\mathrm{rot}$ for our observing epoch is $0.520662(2)$ days. 

It is interesting to note that while \citet{mikulasek2011} reported a near-sinusoidal variation of the rotation period of CU\,Vir with a timescale of $\approx 36$ years using multi-waveband data (including radio), \citet{ravi2010} could detect no period changes in CU\,Vir by performing a timing analysis with the radio pulses observed at epochs separated by $\approx 10$ years (1998--2008). The period obtained from their timing analysis was $0.52071721(9)$, which is longer than the $P_\mathrm{rot}$ for our observing epoch as predicted by the ephemeris of \citet{mikulasek2011}. 
We find that if we use the rotation period of \citet{ravi2010} to phase the lightcurves, the pulses at latter epochs now arrive earlier than those at earlier epochs. This suggests that the true rotation period at our observing epoch is shorter than that estimated by \citet{ravi2010}, but longer than that predicted by the ephemeris of \citet{mikulasek2011}.

In order to estimate the true rotation period for our observations, we used the average profiles for the leading and trailing pulses (\S\ref{subsubsec:average_profile}) to cross-correlate with the pulses at different epochs at respective frequencies (1332.5 MHz for the leading and 2356.5 MHz for the trailing pulses). The lags corresponding to the highest correlation strengths were used to infer the arrival times. The width of the cross-correlation function within which the correlation strength decreases by 5\% from its maximum value is assigned as the uncertainty in the arrival times.

Along with the data from our monitoring campaign, we also used the data reported by \citet{das2021}. In their campaign, the star was observed on 2019--06--12 and 2019--07--23 with the VLA over 1--4 GHz. On 2019--06--12, both pulses were covered, whereas on the latter day, only of the pulses was covered. We chose the earlier epoch of observation (2019--06--12). For the leading and trailing pulses, we used the measurements corresponding to the top and second panels of Figure 2 of \citet{das2021} respectively. Although the corresponding frequencies (1.5 and 2.6 GHz) are not identical to those used for the ATCA data, the error incurred in the arrival times is smaller than the uncertainties obtained from the cross-correlation method described in the preceding paragraph\footnote{\citet{trigilio2008} reported a lag of 14--16 minutes between pulses at 1.4 and 2.5 GHz. Assuming a linear relation between lag and difference in frequencies \citep[approximately valid for small bandwidth, ][]{das2020a0}, the expected lag between pulses at 1.50 and 1.33 GHz and that between 2.55 and 2.36 GHz are $\sim 2$ minutes in each case. This is smaller than the uncertainties in the times of arrival of the pulses, which are $\approx 5$ minutes.}.

In the top panel of Figure \ref{fig:arrival_phase_vs_epoch}, the relative arrival phases, calculated using the ephemeris of \citet{mikulasek2011} are shown for the pulses at different epochs. For both leading and trailing pulses, the arrival phases increases linearly over the course of our observing session. We fitted straight lines to the data for both leading and trailing pulses and the slopes came out to be $(4.6\pm0.4)\times10^{-5}$ cycles/day (leading) and $(4.6\pm0.2)\times10^{-5}$ cycles/day (trailing).


We next varied the rotation period between 0.520662 days \citep[the mean period for the rotation period model of ][]{mikulasek2011} and 0.52071721 days \citep[the rotation period reported by ][]{ravi2010}, and for each case, calculated the slope of the straight line fitted between arrival phase and epoch of observation. The rotation period that minimizes the slope came out to be 
$0.52068827(8)$ days (using the data for the leading pulse) and $0.52068815(6)$ days (using the data for the trailing pulse), with a mean of $P_0=0.5206882$ days. Note that excluding the data from 2019 results in a period of $0.520689(4)$ days (leading pulse) and $0.520683(3)$ days (trailing pulse), with a mean of $0.520686$ days. 
In the bottom panel of Figure \ref{fig:arrival_phase_vs_epoch}, the arrival phases are shown that are calculated using $P_0=0.5206882$ days as the rotation period. As expected, no systematic trends are observed in this case. Thus, with these radio data alone, we do not have any evidence of a period evolution between 2019 and 2024.


The rotation period obtained from the radio data deviates from that predicted by the ephemeris of \citet{mikulasek2011} by more than $15\sigma$ (taking into account the uncertainties in the various parameters of the ephemeris). However, it should be noted that the ephemeris of \citet{mikulasek2011} was derived using data taken between 1949 and 2011, and thus the period in 2024 was obtained by extrapolating the phase function, which need not be valid. This limitation motivates a more extensive timing exercise (covering a much longer timeline), which will be the subject of a future publication.

To summarize, there is strong evidence that CU\,Vir has undergone a spin-up between 2008 and 2024 (from a period of 0.5207172 days to 0.5206882 days), qualitatively consistent with the prediction of \citet{mikulasek2011}, though at a rate smaller than that predicted by this model. 

\section{Summary}\label{sec:summary}
As the first discovered MRP, CU\,Vir is the most extensively studied object in the current MRP sample. The spectral properties of its radio pulses have been studied in detail \citep{trigilio2011,das2021}, but the stability of those properties has remained to be investigated until now. This monitoring campaign, involving observation of its radio pulses at 36 epochs using the ATCA at 1--3 GHz, is the first step towards overcoming this limitation.
The key points that emerged are listed below:
\begin{enumerate}
    \item Similar to pulses from pulsars, pulses from CU\,Vir also exhibit phase-jitter, which refers to the phenomenon in which pulses at consecutive epochs arrive at slightly different phase-windows. The observed phase-jitter is as large as 0.014 rotation cycle that translates to $\approx 10$ minutes for CU\,Vir. This phase-jitter should be incorporated in the uncertainty in the pulse arrival times (important for future timing experiments).
    \item Pulse fluence and peak flux densities are strongly correlated so that one can be used as a proxy for the other.
    \item Although past studies indicated that only one of the pulses (leading pulse) is intermittent at 13 cm, we find that the trailing pulses can also `disappear' at that frequency though the probability is much less. With the help of continuous frequency coverage, we find that the apparent `intermittent' nature is actually caused by inherent spectral shapes, broadband epoch to epoch variation of flux densities, and telescope sensitivity.
    \item The fluctuations (quantified by variability indices) exhibited by the trailing pulse monotonically increases from $\approx 25\%$ to $\approx 40\%$ with increasing frequency. The leading pulse exhibits higher fluctuations than that for the trailing pulse at all frequencies, and also exhibits an opposite trend in the spectral variation, in which the variability index decreases from $\approx 60\%$ to $\approx 40\%$ with increasing frequency. This difference observed between leading and trailing pulses is difficult to explain by external phenomena such as interstellar scintillation, or considering random instabilities at the emission sites alone.
    \item A scenario that could explain the observed spectral dependence of variability indices is one involving CBOs (confined to magnetic equatorial regions), randomly affecting the emission, but to equal extents at all frequencies; and frequency dependent instabilities at the emission sites.
    \item The observed spectral and temporal variations do not rule out the existence of a characteristic spectral shapes of ECME for CU\,Vir. For our sample spectra, the average spectrum for the leading pulse can be described by a broken power law, with spectral indices of 0.65 and 3.5 below and above a break frequency of 2 GHz respectively. For the trailing pulse, the corresponding spectrum can be approximated with a flat spectrum up to 2.4 GHz, beyond that, the flux density declines with a spectral index of 5.8.
    \item We were able to average over 15 pulses for the leading and 13 pulses for the trailing pulse to construct average profiles. By using constraints from our sample, we find that $\sim 30$ pulses are needed to extract global pulse properties for either leading or trailing pulses. 
    \item CU\,Vir has spun-up between 2008--2024, but at a slower rate than that predicted by the ephemeris of \citet{mikulasek2011}. By combining our data with those reported by \citet{das2021}, the best-fit rotation period is found to be 0.5206882 days.
\end{enumerate}

Thus, our pilot study has demonstrated the usefulness of such a campaign. However, to pinpoint the origin of temporal evolution, difference between pulses produced by same magnetic hemispheres, but observed at different phases, and also to extract global properties, it will be important to conduct more extensive campaigns covering wider frequency ranges at higher number of epochs and more MRPs.

\begin{acknowledgments}
We thank the referee for their constructive criticism that helped us to significantly improve the manuscript.
The Australia Telescope Compact Array is part of the Australia Telescope National Facility (\url{https://ror.org/05qajvd42}) which is funded by the Australian Government for operation as a National Facility managed by CSIRO.
We acknowledge the Gomeroi people as the Traditional Owners of the Observatory site.
This research has made use of NASA's Astrophysics Data System.
\end{acknowledgments}

\begin{contribution}

BD conceived the idea of a monitoring campaign of CU\,Vir, led the observing proposal, data analysis, interpretation and preparation of the manuscript. 
HB, AZ and JP helped in data acquisition, reduction and interpretation.
PC contributed to the observing proposal and data interpretation.
JM, AG, BB and GH contributed to data interpretation.
All authors contributed to the manuscript preparation.

\end{contribution}

%
\facilities{ATCA}

\software{CASA \citep{mcmullin2007},
            numpy \citep{harris2020array},
            astropy \citep{2013A&A...558A..33A,2018AJ....156..123A,2022ApJ...935..167A}
          }


\appendix

\section{Observation epochs, individual Lightcurves and peak flux density spectra}

\setcounter{figure}{0}
\renewcommand{\thefigure}{\thesection\arabic{figure}}

\setcounter{table}{0}
\renewcommand{\thetable}{\thesection\arabic{table}}

\begin{deluxetable}{lcc}
\tablecaption{ATCA observation of CU\,Vir between March 31, 2024 and September 9, 2024. The rotational phases are calculated using the ephemeris of \citet{mikulasek2011}.\label{tab:observing_epochs}}
\tablehead{
\colhead{Date} &
\colhead{MJD range} &
\colhead{Rotational phase range}
}
\startdata
{2024--4--01} & $60401.718\pm0.054$ & $0.339\pm0.105$ \\
{2024--4--02} & $60402.763\pm0.046$ & $0.345\pm0.088$ \\
{2024--4--03} & $60403.802\pm0.047$ & $0.34\pm0.089$ \\
{2024--4--19} & $60419.628\pm0.078$ & $0.738\pm0.151$ \\
{2024--4--28} & $60428.459\pm0.048$ & $0.699\pm0.092$ \\
{2024--4--29} & $60429.498\pm0.053$ & $0.695\pm0.102$ \\
{2024--4--30} & $60430.539\pm0.051$ & $0.694\pm0.099$ \\
{2024--5--01} & $60431.568\pm0.065$ & $0.67\pm0.124$ \\
{2024--5--02} & $60432.613\pm0.063$ & $0.677\pm0.12$ \\
{2024--6--07} & $60468.358\pm0.049$ & $0.325\pm0.095$ \\
{2024--6--09} & $60470.439\pm0.055$ & $0.322\pm0.105$ \\
{2024--6--10} & $60471.486\pm0.052$ & $0.333\pm0.101$ \\
{2024--6--12} & $60473.566\pm0.056$ & $0.328\pm0.108$ \\
{2024--7--01} & $60492.504\pm0.059$ & $0.699\pm0.113$ \\
{2024--7--02} & $60493.528\pm0.074$ & $0.664\pm0.142$ \\
{2024--7--07} & $60498.247\pm0.071$ & $0.728\pm0.135$ \\
{2024--7--12} & $60503.461\pm0.039$ & $0.741\pm0.074$ \\
{2024--7--13} & $60504.481\pm0.054$ & $0.699\pm0.105$ \\
{2024--7--19} & $60510.215\pm0.054$ & $0.711\pm0.104$ \\
{2024--7--21} & $60512.305\pm0.047$ & $0.725\pm0.091$ \\
{2024--7--23} & $60514.384\pm0.051$ & $0.718\pm0.097$ \\
{2024--7--24} & $60515.422\pm0.055$ & $0.711\pm0.106$ \\
{2024--8--16} & $60538.133\pm0.053$ & $0.326\pm0.103$ \\
{2024--8--17} & $60539.174\pm0.055$ & $0.326\pm0.105$ \\
{2024--8--18} & $60540.218\pm0.05$ & $0.331\pm0.095$ \\
{2024--8--29} & $60551.155\pm0.05$ & $0.336\pm0.097$ \\
{2024--8--30} & $60552.203\pm0.047$ & $0.348\pm0.089$ \\
{2024--8--31} & $60553.236\pm0.052$ & $0.332\pm0.101$ \\
{2024--9--01} & $60554.281\pm0.055$ & $0.338\pm0.105$ \\
{2024--9--02} & $60555.319\pm0.053$ & $0.332\pm0.102$ \\
{2024--9--09} & $60562.087\pm0.054$ & $0.33\pm0.103$ \\
{2024--9--11} & $60564.196\pm0.078$ & $0.38\pm0.151$ \\
{2024--9--21} & $60574.266\pm0.043$ & $0.721\pm0.083$ \\
{2024--9--22} & $60575.304\pm0.048$ & $0.713\pm0.092$ \\
{2024--9--27} & $60581.039\pm0.041$ & $0.728\pm0.079$ \\
{2024--9--29} & $60582.065\pm0.057$ & $0.699\pm0.109$ \\
\enddata
\end{deluxetable}

\begin{figure*}
    \centering
    \includegraphics[width=0.2\textwidth]{  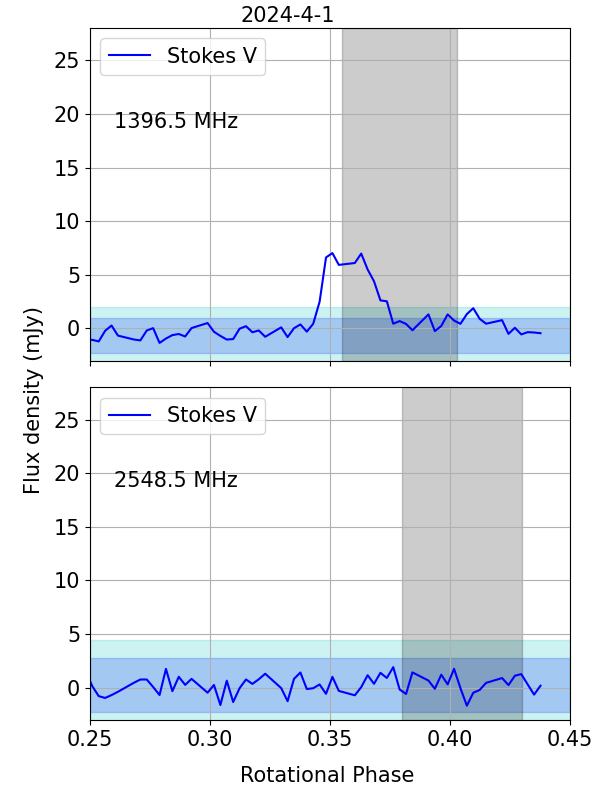}
    \includegraphics[width=0.2\textwidth]{  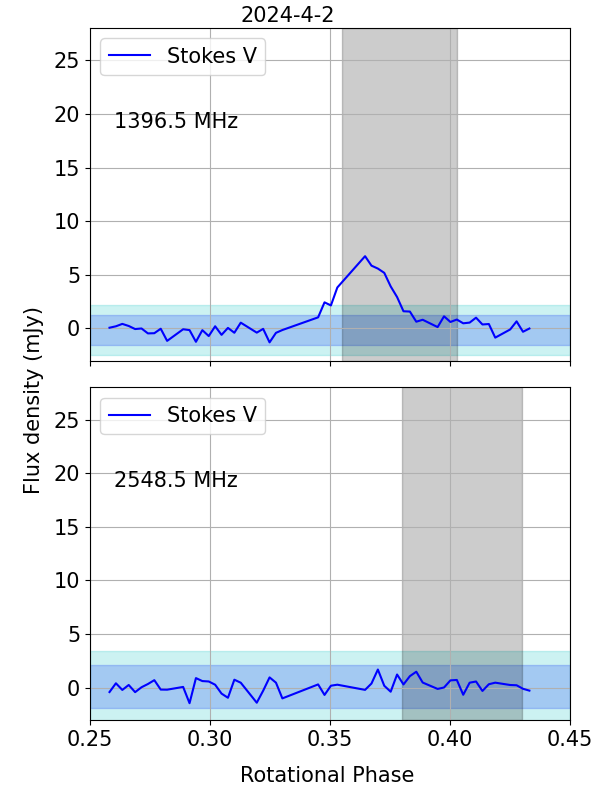}
    \includegraphics[width=0.2\textwidth]{  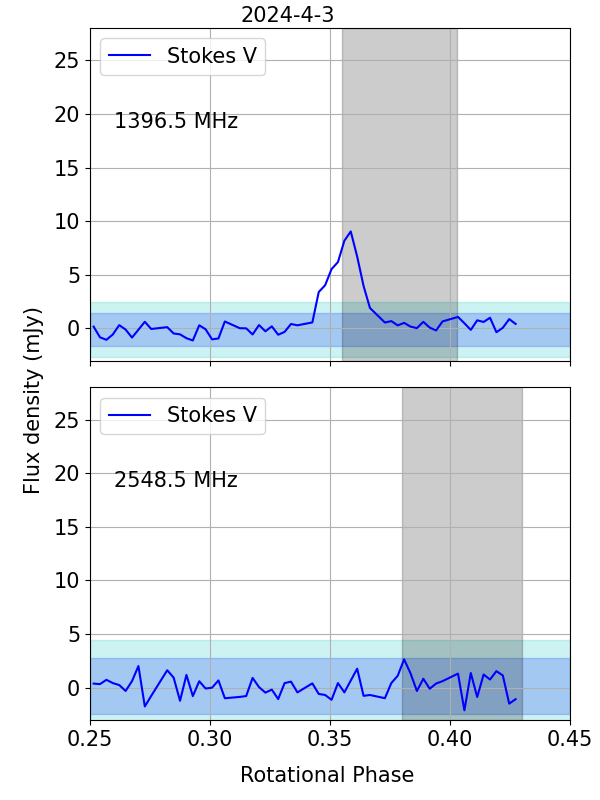}
    \includegraphics[width=0.2\textwidth]{  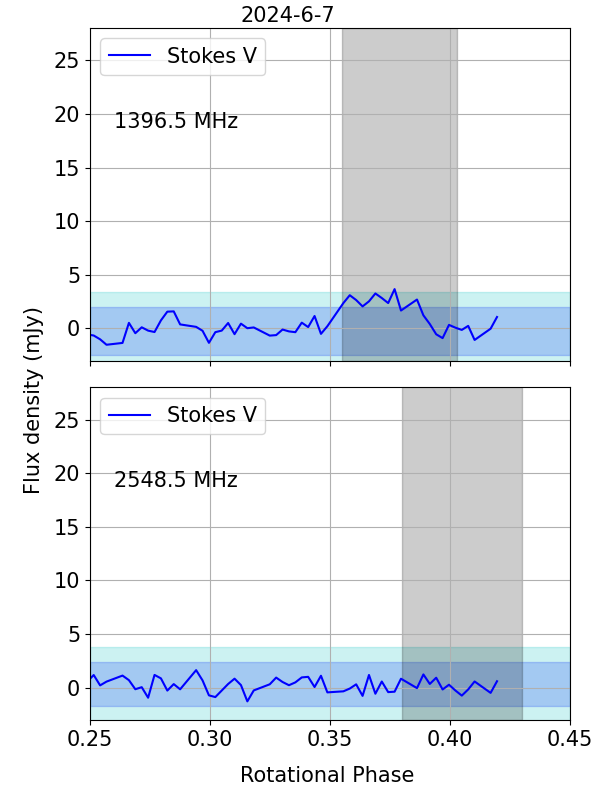}
    \includegraphics[width=0.2\textwidth]{  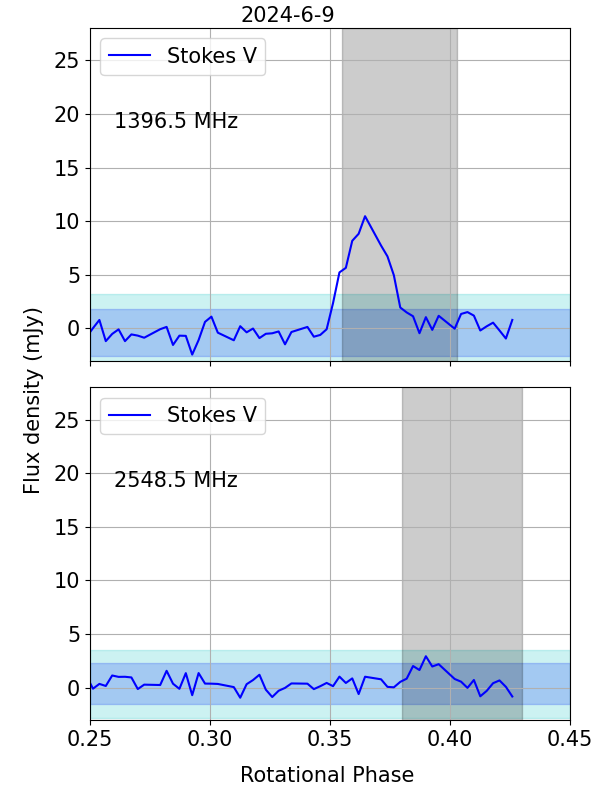}
    \includegraphics[width=0.2\textwidth]{  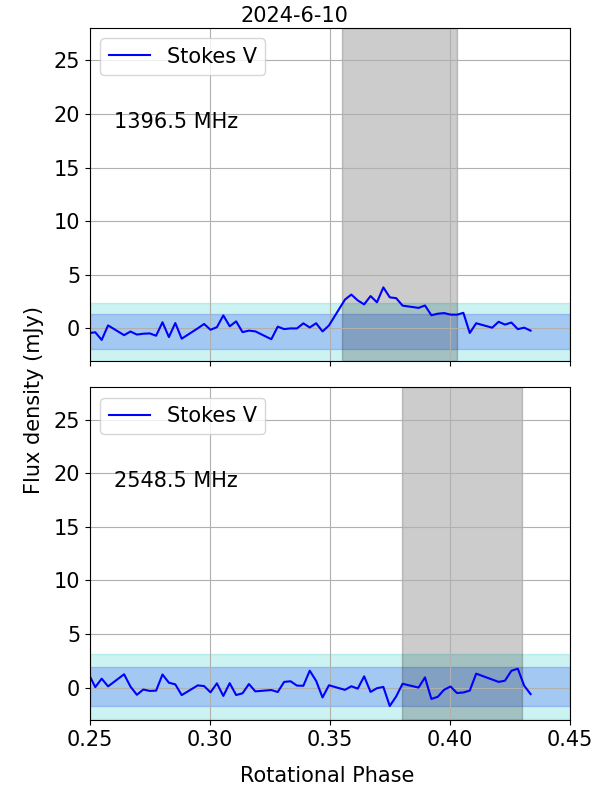}
    \includegraphics[width=0.2\textwidth]{  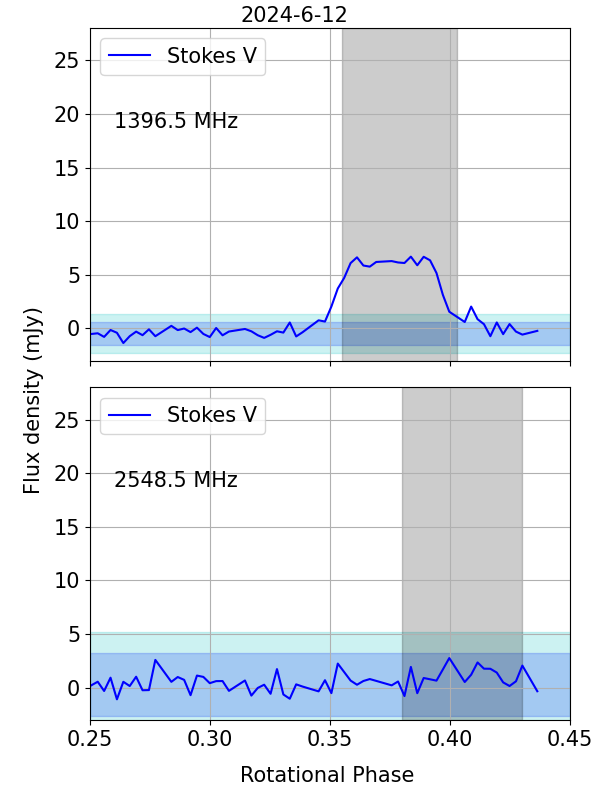}
    \includegraphics[width=0.2\textwidth]{  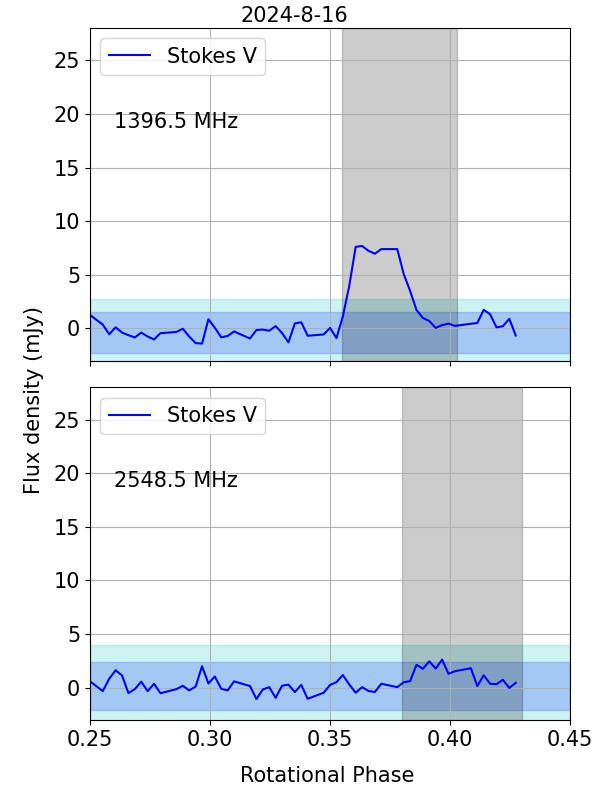}
    \includegraphics[width=0.2\textwidth]{  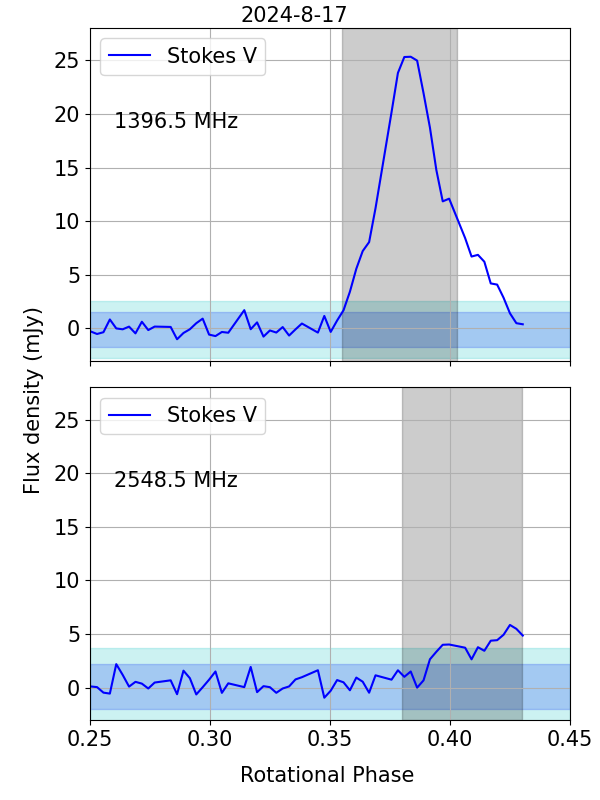}
    \includegraphics[width=0.2\textwidth]{  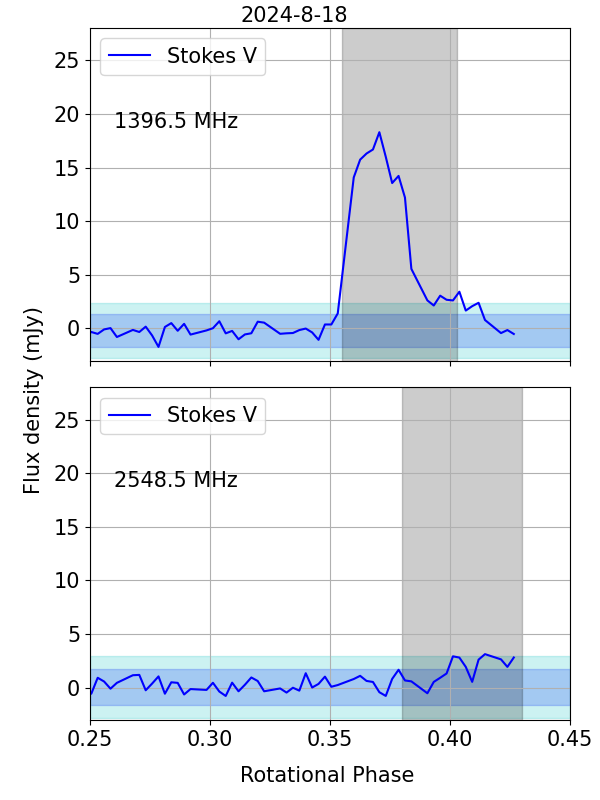}
    \includegraphics[width=0.2\textwidth]{  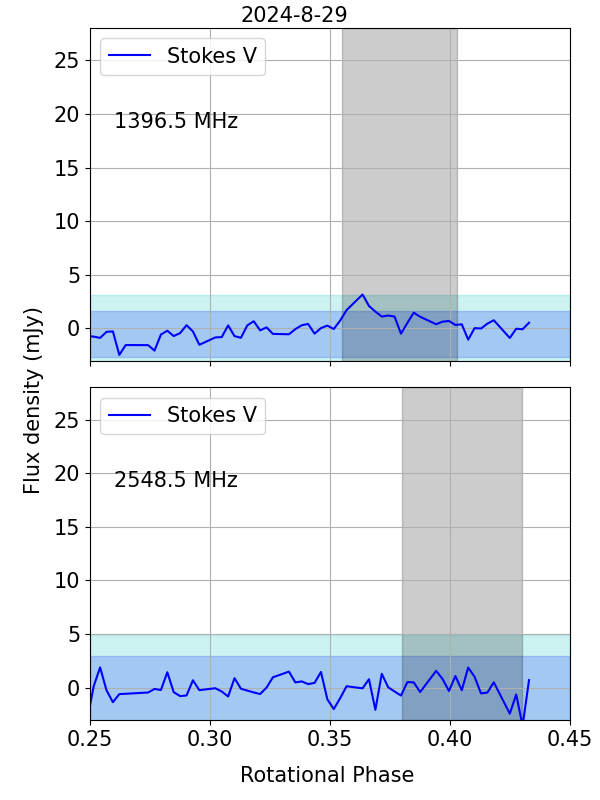}
    \includegraphics[width=0.2\textwidth]{  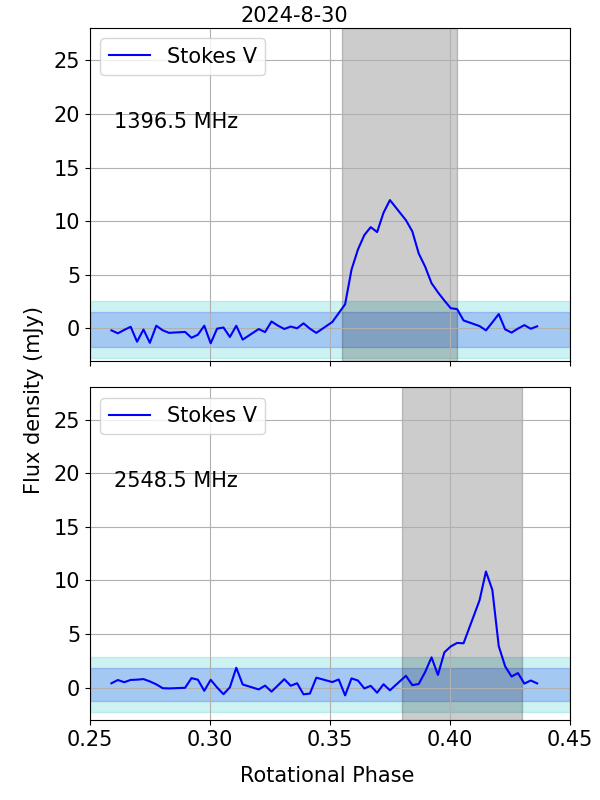}
    \includegraphics[width=0.2\textwidth]{  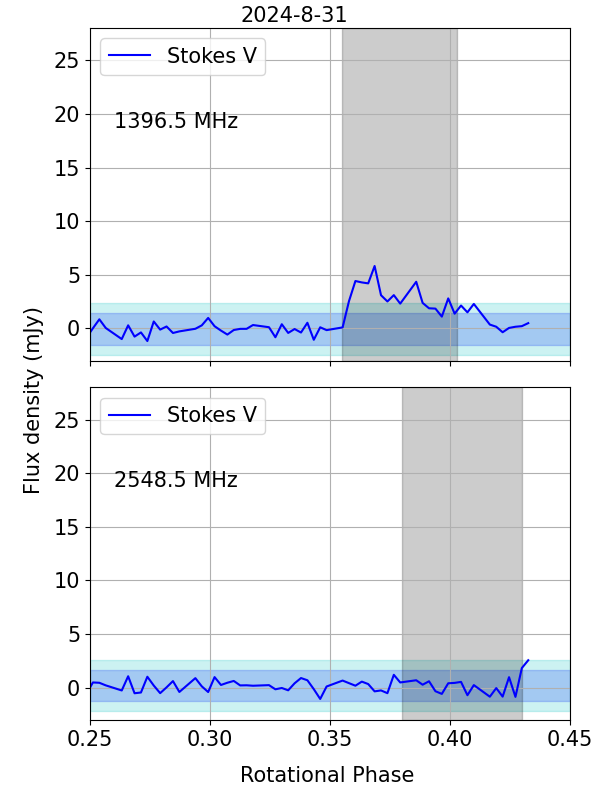}
    \includegraphics[width=0.2\textwidth]{  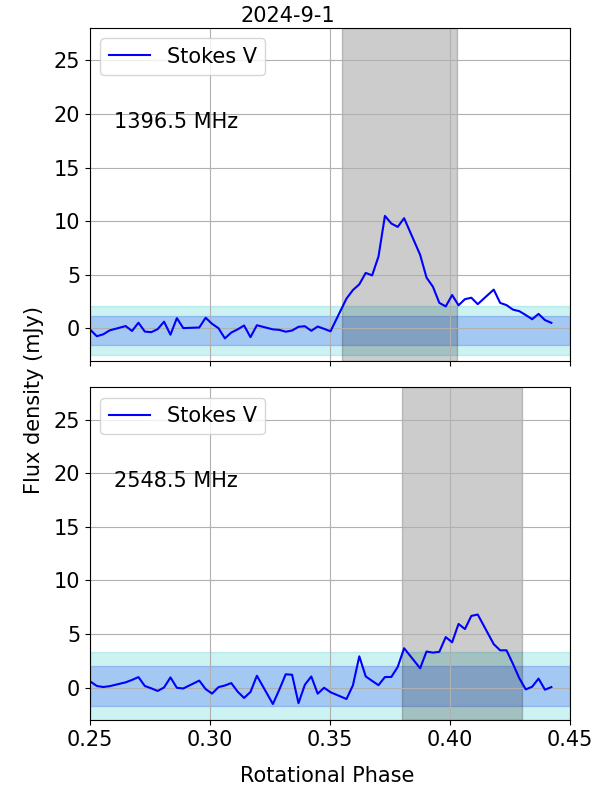}
    \includegraphics[width=0.2\textwidth]{  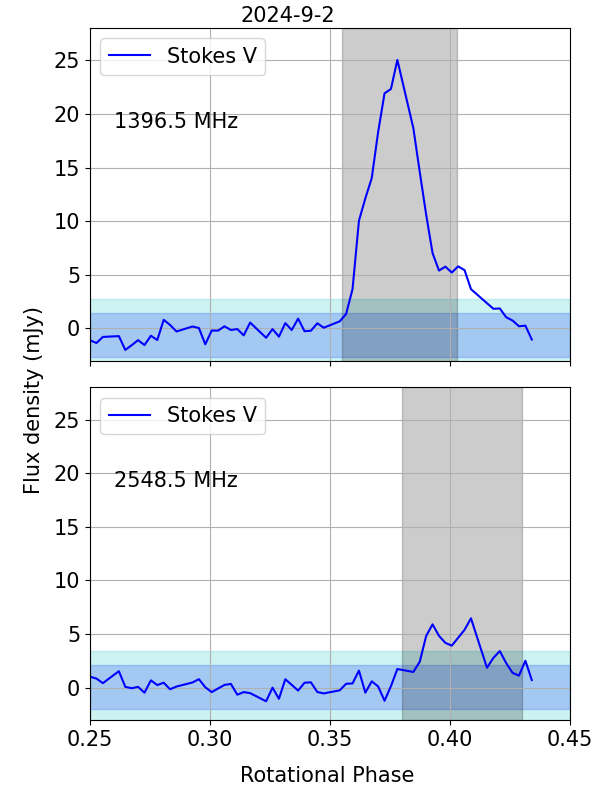}
    \includegraphics[width=0.2\textwidth]{  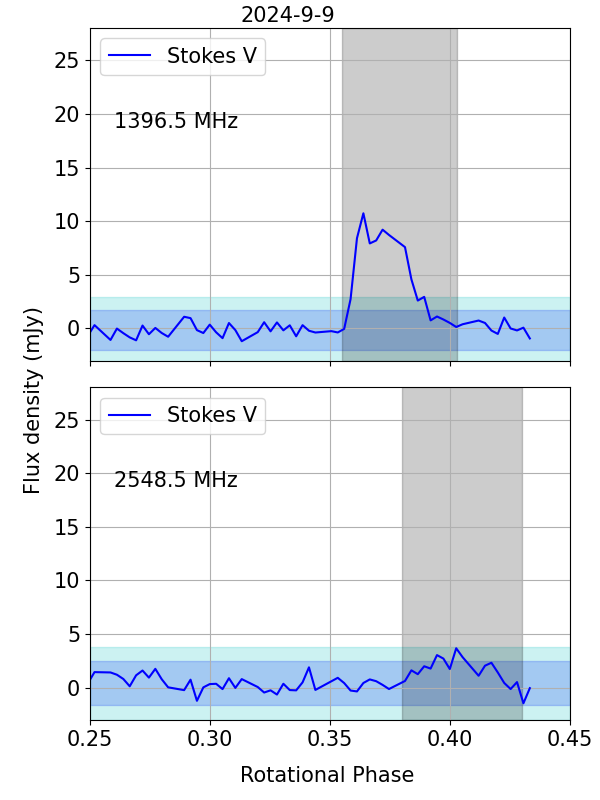}
    \includegraphics[width=0.2\textwidth]{  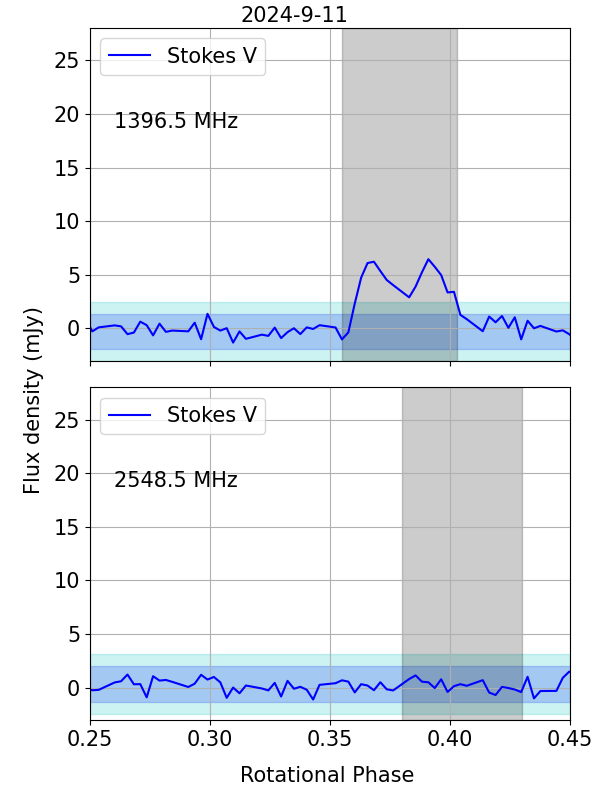}
    \caption{The lightcurves for the leading pulse. The blue and cyan shaded regions correspond to $3\sigma$ and $5\sigma$ for the Stokes V lightcurves. 
    The grey shaded regions approximately mark the rotational phases of arrival of the pulses.
    \label{fig:lc_leading}}
\end{figure*}

\begin{figure*}
    \centering
    \includegraphics[width=0.2\textwidth]{  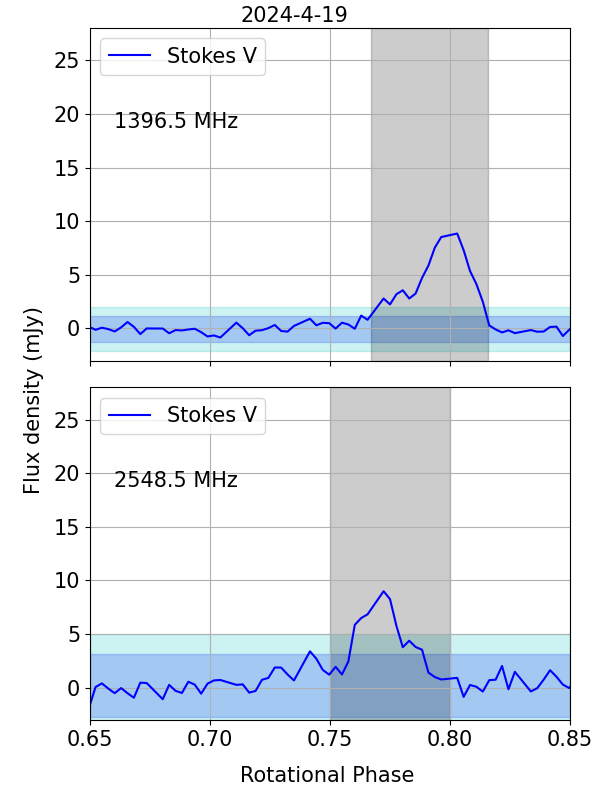}
    \includegraphics[width=0.2\textwidth]{  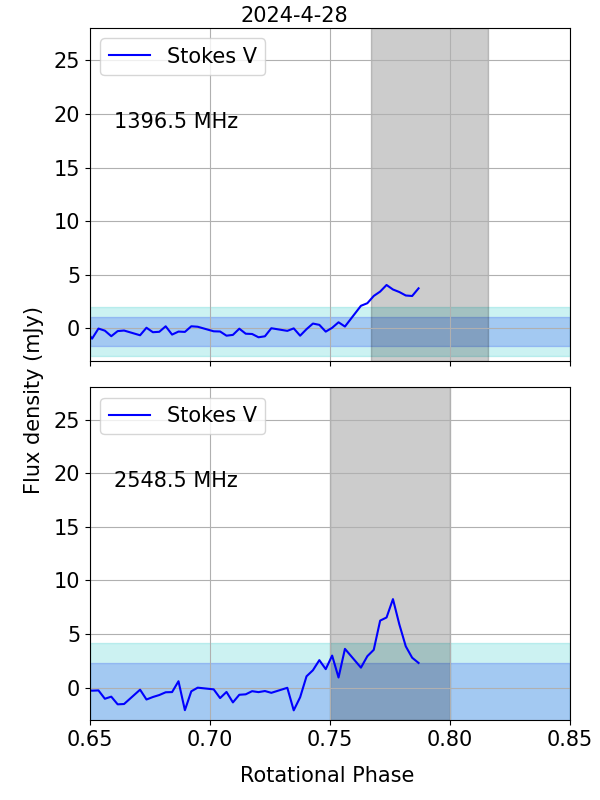}
    \includegraphics[width=0.2\textwidth]{  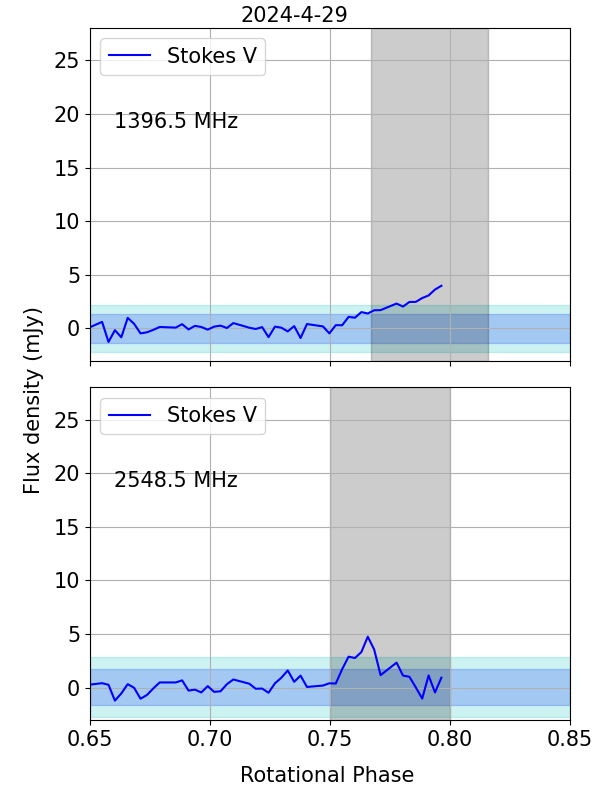}
    \includegraphics[width=0.2\textwidth]{  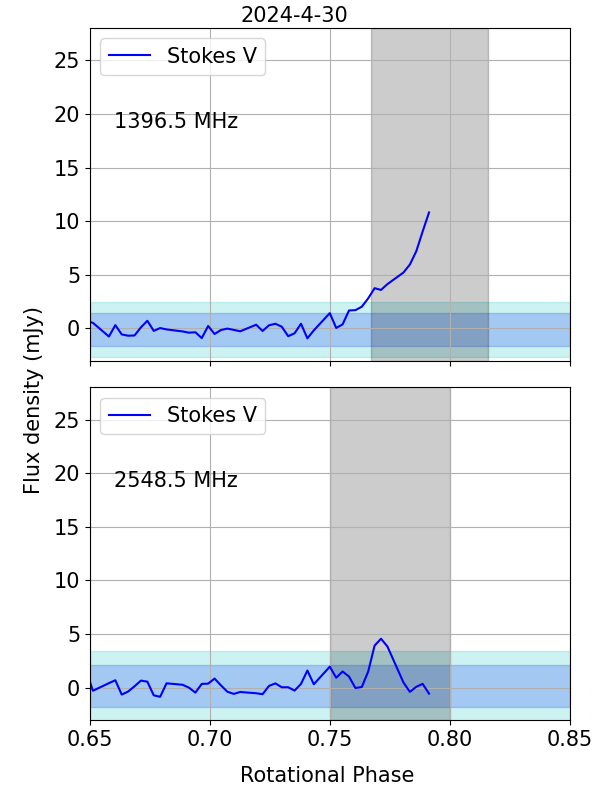}
    \includegraphics[width=0.2\textwidth]{  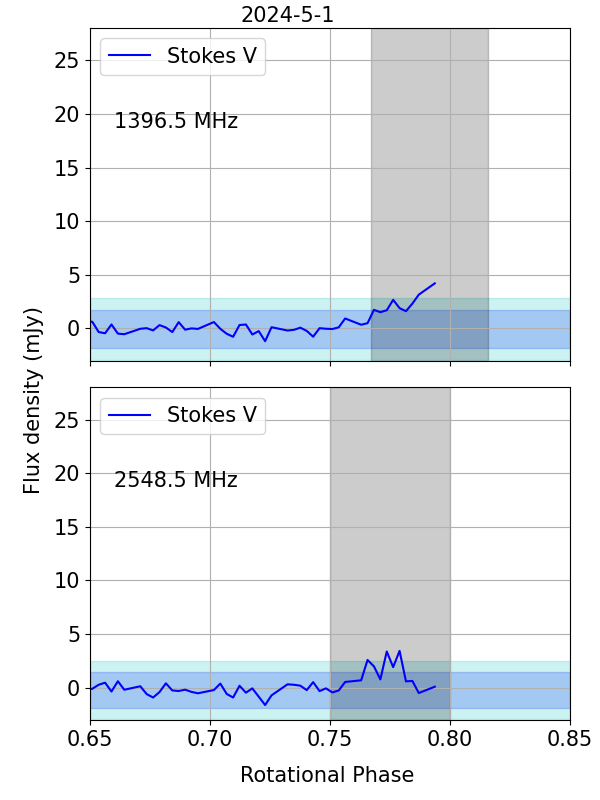}
    \includegraphics[width=0.2\textwidth]{  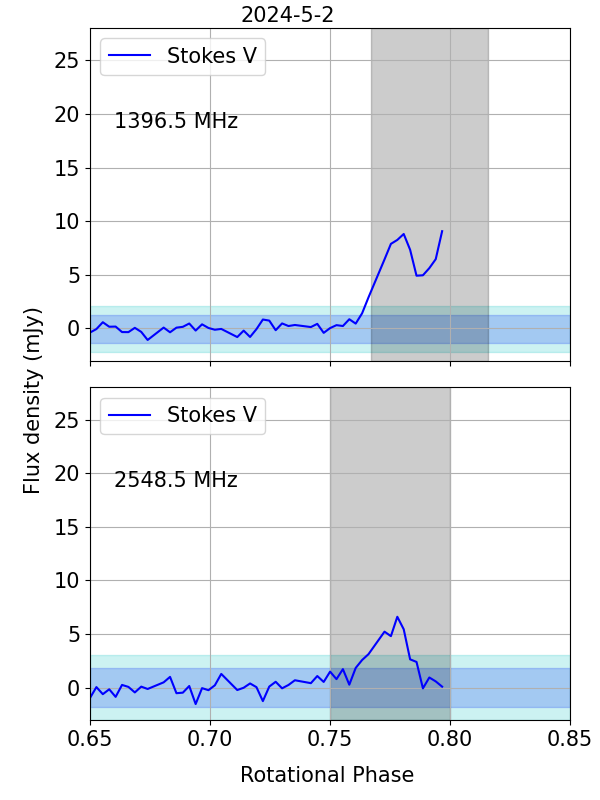}
    \includegraphics[width=0.2\textwidth]{  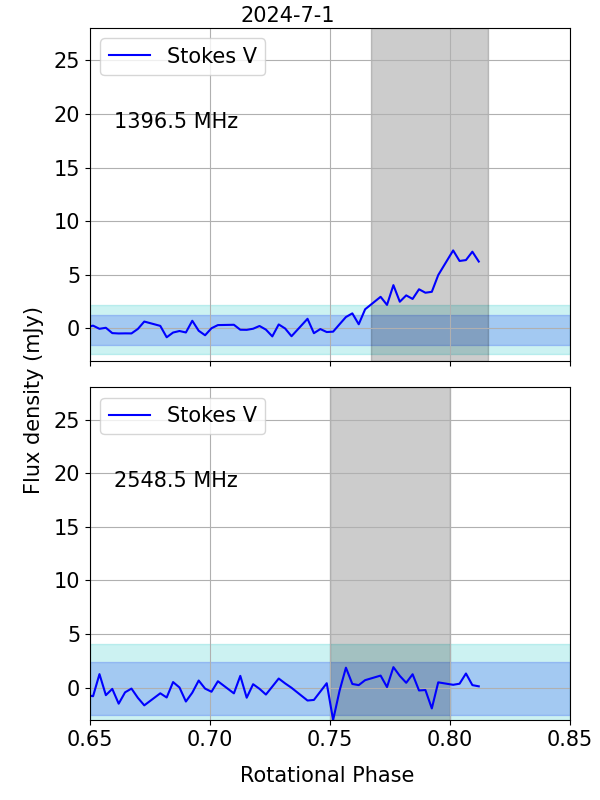}
    \includegraphics[width=0.2\textwidth]{  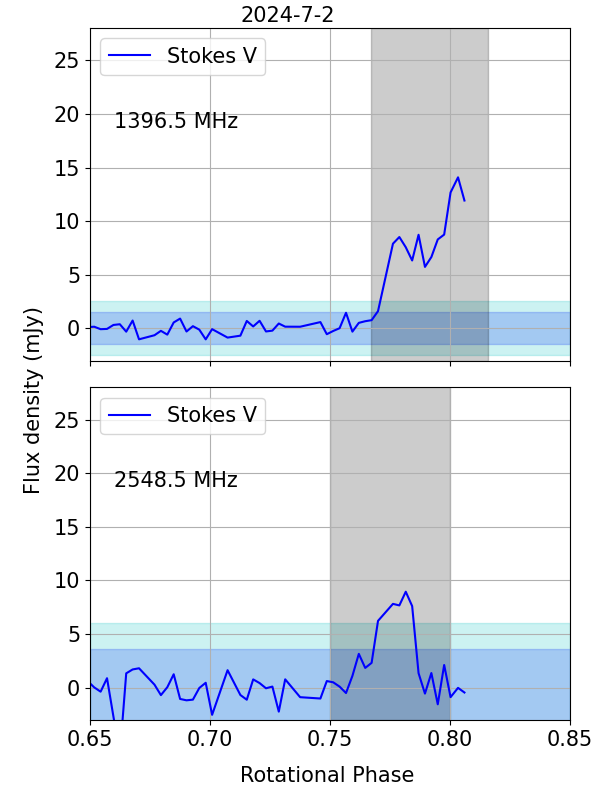}
    \includegraphics[width=0.2\textwidth]{  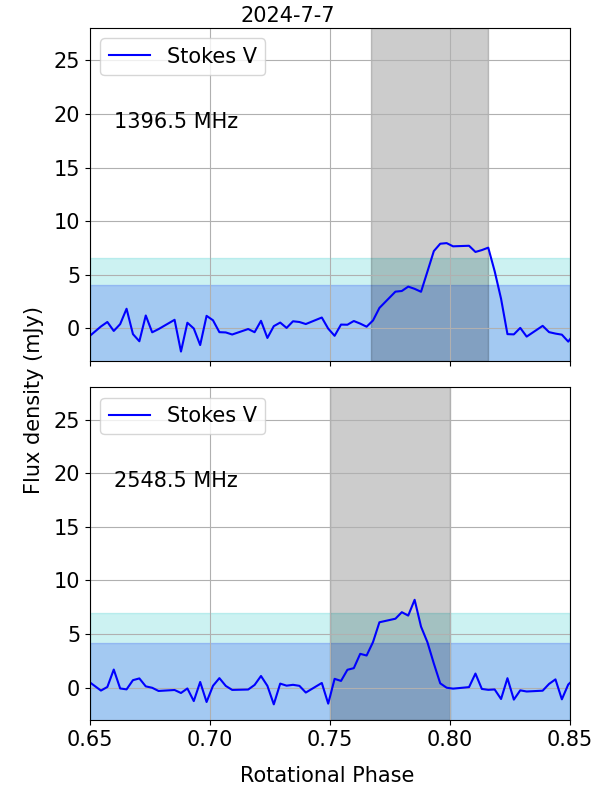}
    \includegraphics[width=0.2\textwidth]{  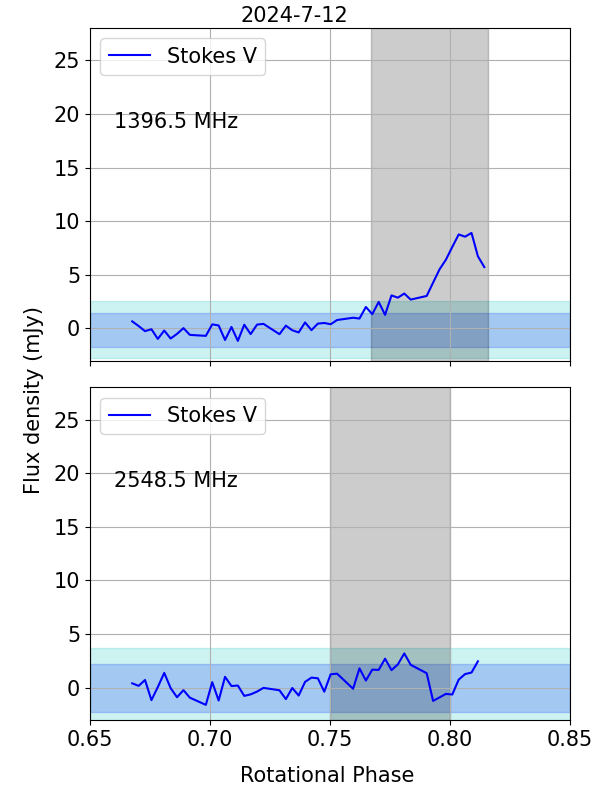}
    \includegraphics[width=0.2\textwidth]{  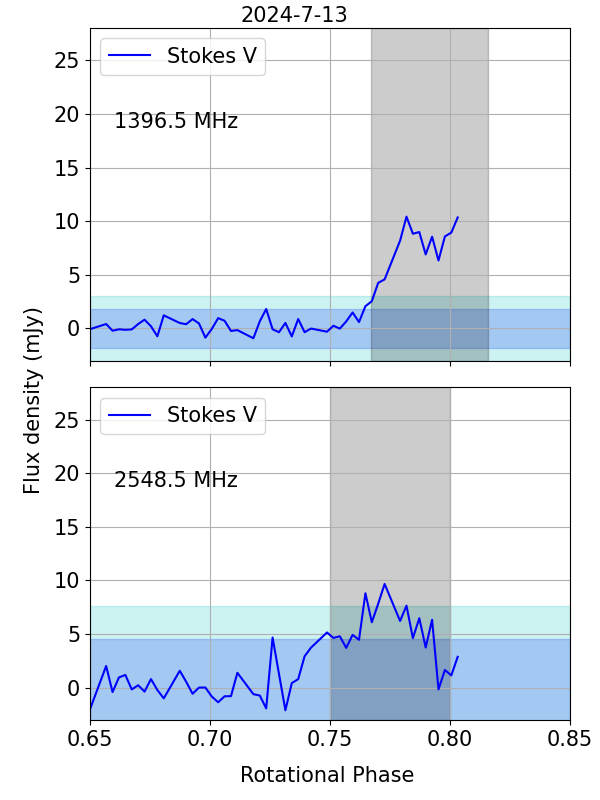}
    \includegraphics[width=0.2\textwidth]{  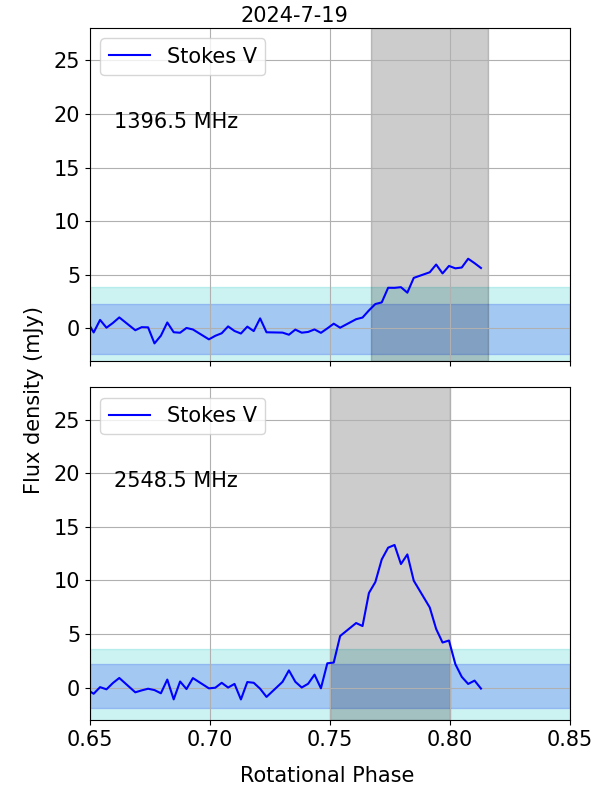}
    \includegraphics[width=0.2\textwidth]{  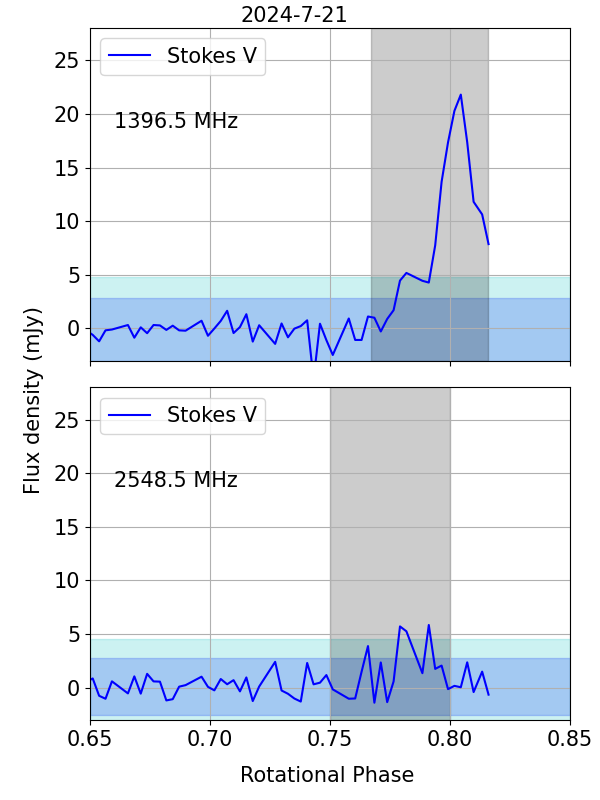}
    \includegraphics[width=0.2\textwidth]{  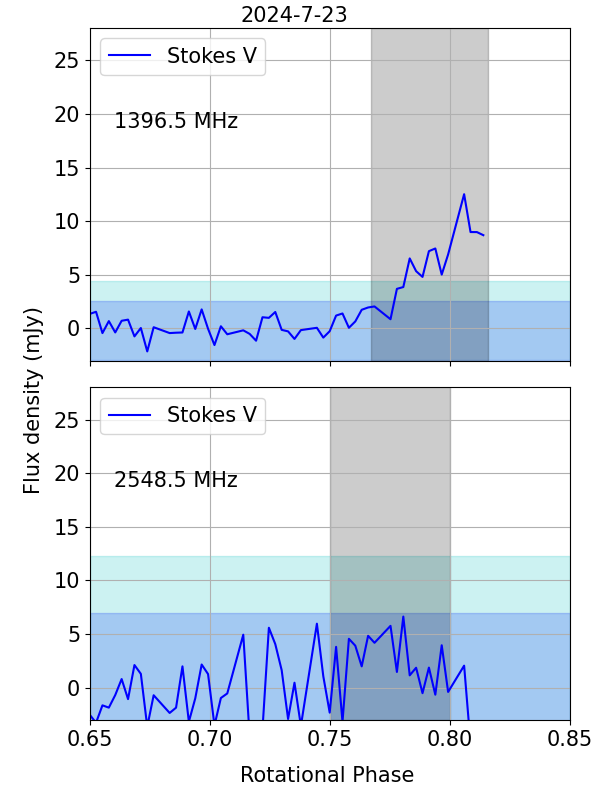}
    \includegraphics[width=0.2\textwidth]{  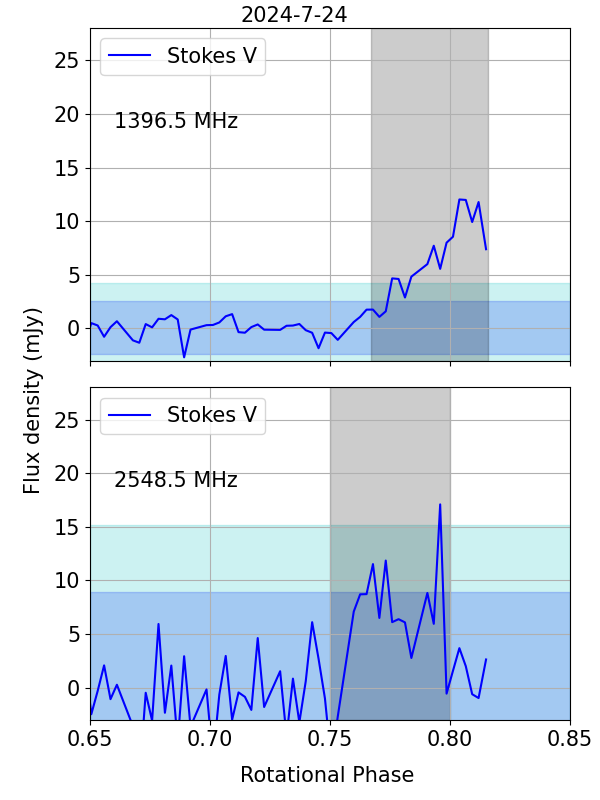}
    \includegraphics[width=0.2\textwidth]{  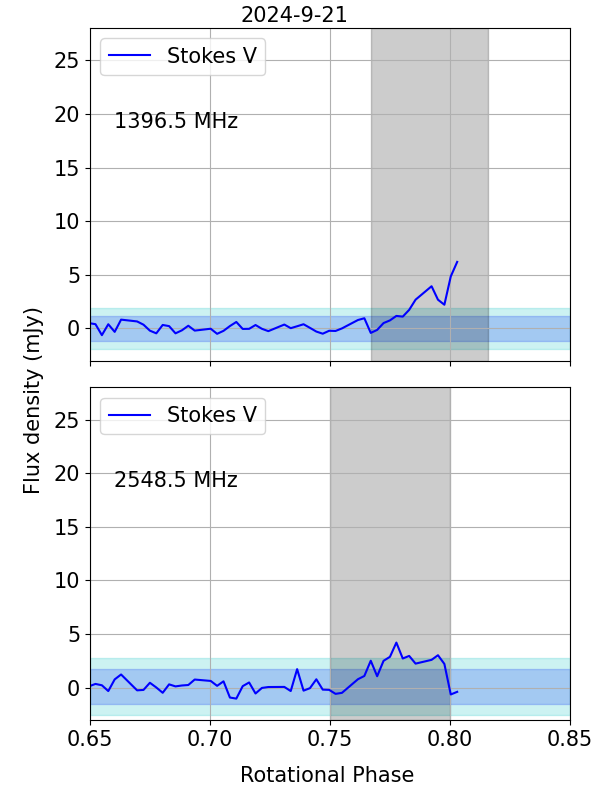}
    \includegraphics[width=0.2\textwidth]{  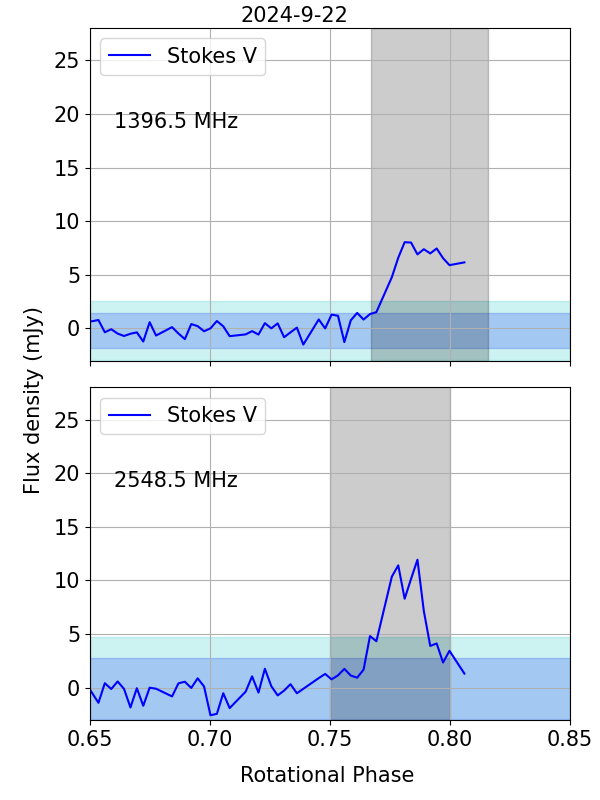}
    \includegraphics[width=0.2\textwidth]{  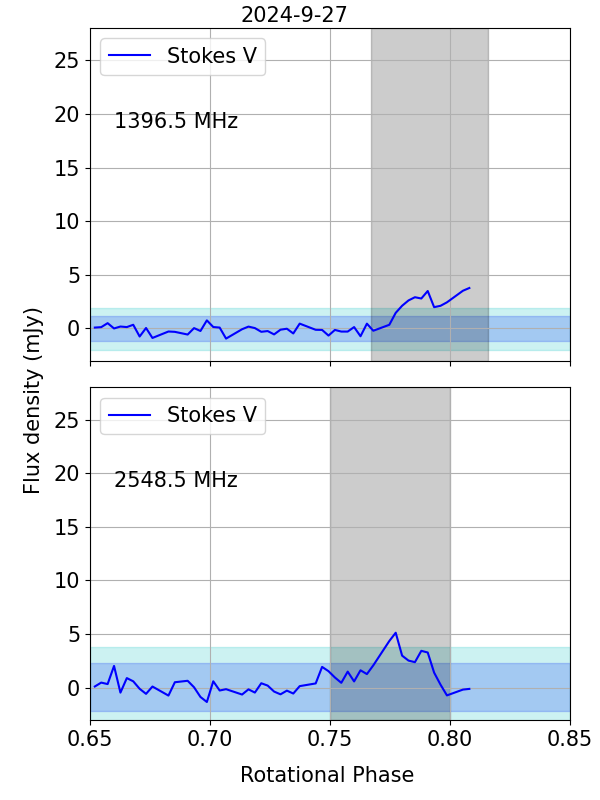}
    \includegraphics[width=0.2\textwidth]{  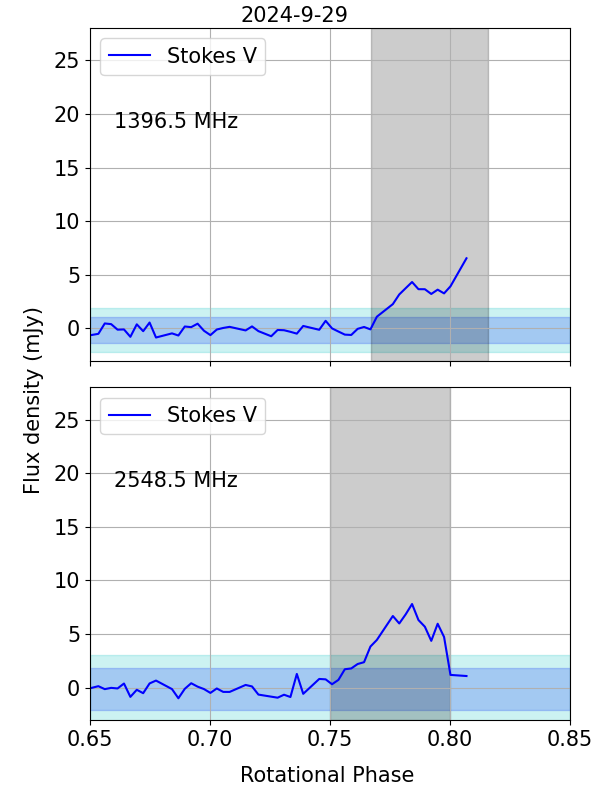}
    \caption{The lightcurves for the trailing pulse. The blue and cyan shaded regions correspond to $3\sigma$ and $5\sigma$ for the Stokes V lightcurves.
    The grey shaded regions approximately mark the rotational phases of arrival of the pulses.\label{fig:lc_trailing}}
\end{figure*}

\begin{figure*}
    \centering
    \includegraphics[trim={0cm 0cm 0cm 1.15cm}, clip, width=0.24\textwidth]{  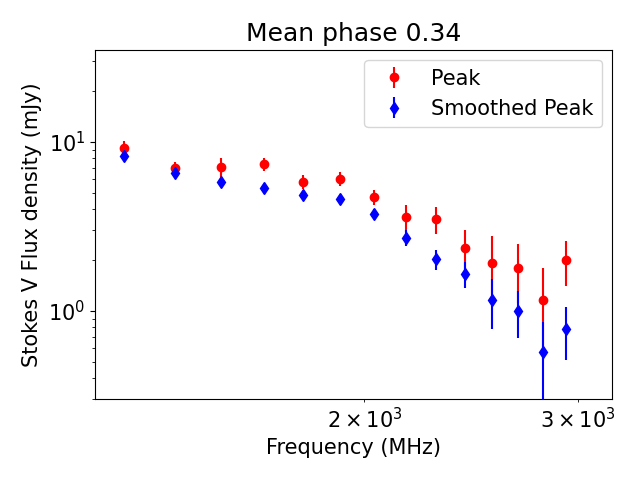}
    \includegraphics[trim={0cm 0cm 0cm 1.15cm}, clip, width=0.24\textwidth]{  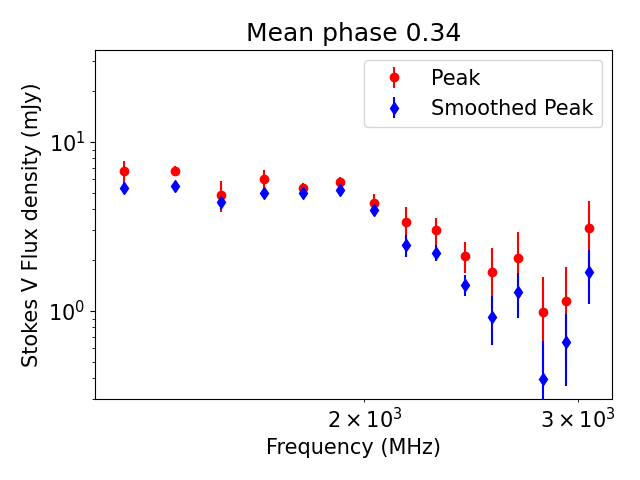}
    \includegraphics[trim={0cm 0cm 0cm 1.15cm}, clip, width=0.24\textwidth]{  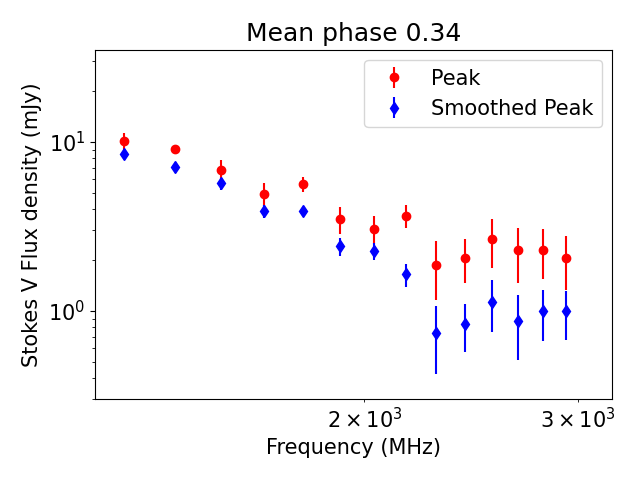}
    \includegraphics[trim={0cm 0cm 0cm 1.15cm}, clip, width=0.24\textwidth]{  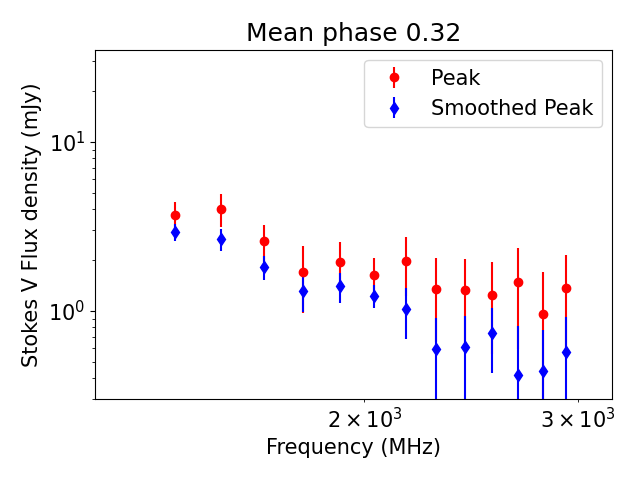}
    \includegraphics[trim={0cm 0cm 0cm 1.15cm}, clip, width=0.24\textwidth]{  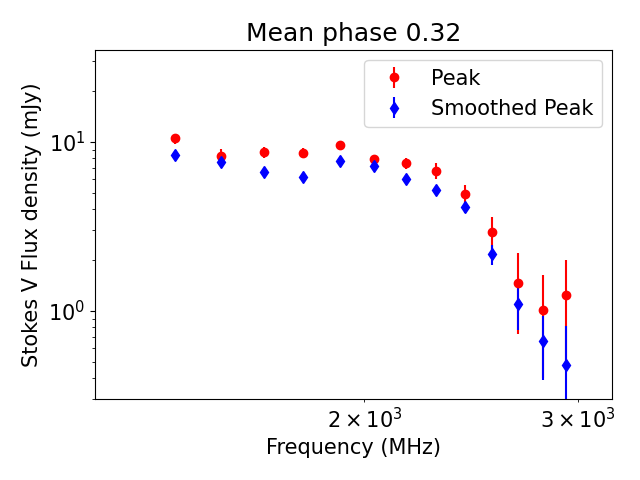}
    \includegraphics[trim={0cm 0cm 0cm 1.15cm}, clip, width=0.24\textwidth]{  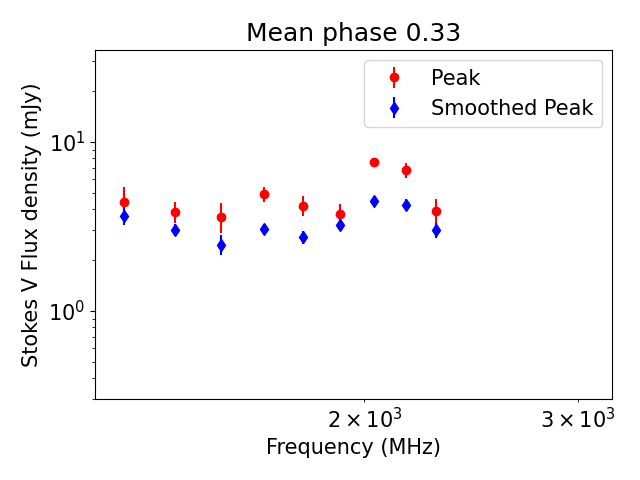}
    \includegraphics[trim={0cm 0cm 0cm 1.15cm}, clip, width=0.24\textwidth]{  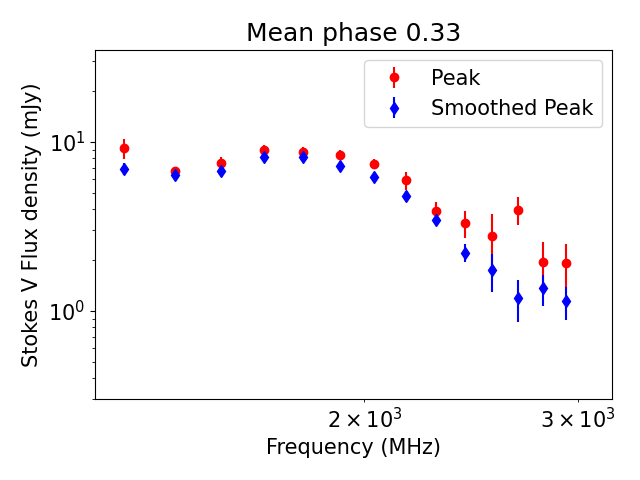}
    \includegraphics[trim={0cm 0cm 0cm 1.15cm}, clip, width=0.24\textwidth]{  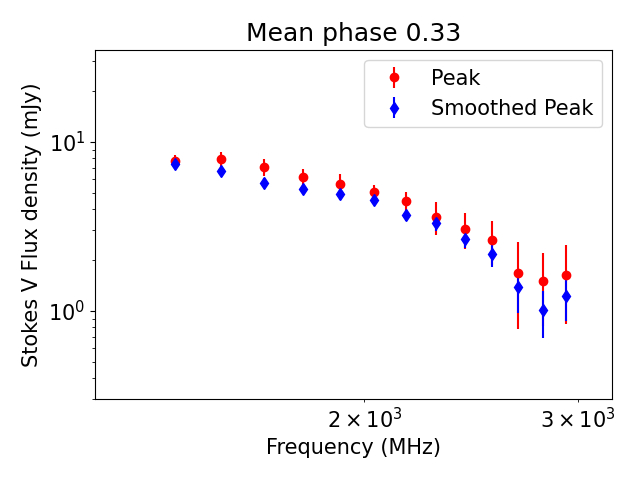}
    \includegraphics[trim={0cm 0cm 0cm 1.15cm}, clip, width=0.24\textwidth]{  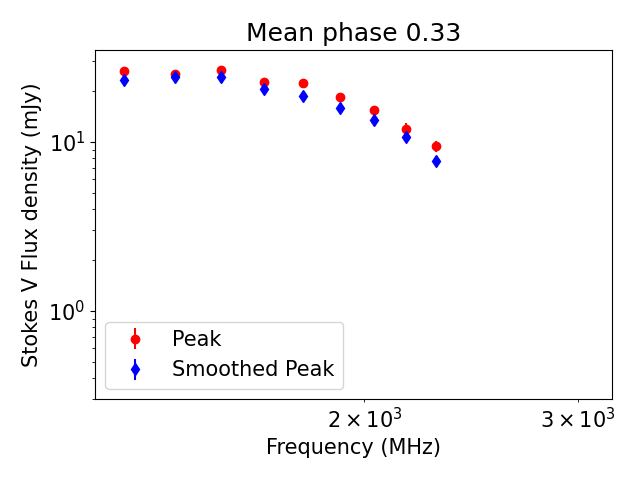}
    \includegraphics[trim={0cm 0cm 0cm 1.15cm}, clip, width=0.24\textwidth]{  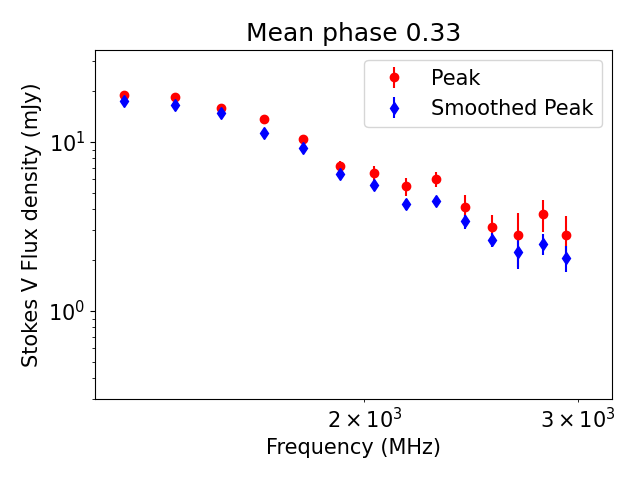}
    \includegraphics[trim={0cm 0cm 0cm 1.15cm}, clip, width=0.24\textwidth]{  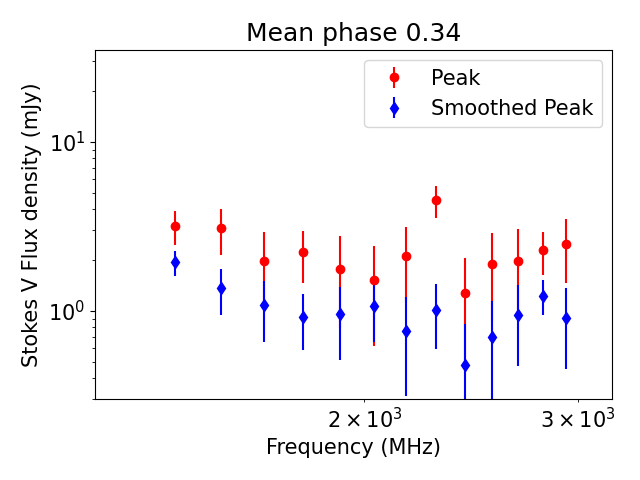}
    \includegraphics[trim={0cm 0cm 0cm 1.15cm}, clip, width=0.24\textwidth]{  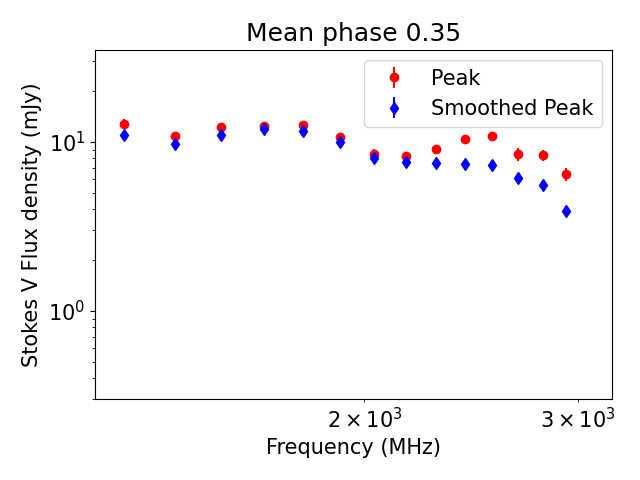}
    \includegraphics[trim={0cm 0cm 0cm 1.15cm}, clip, width=0.24\textwidth]{  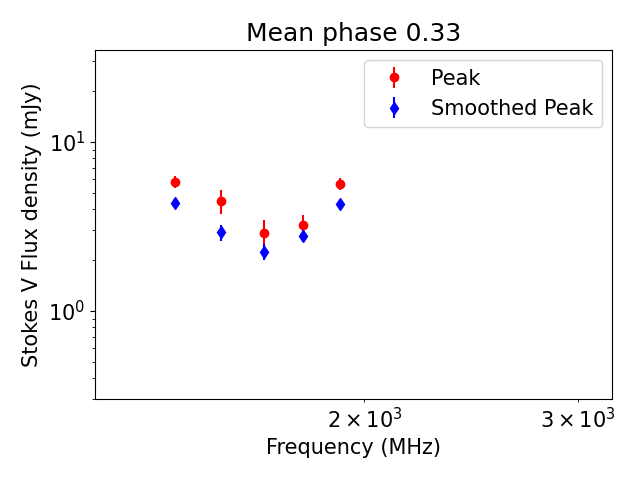}
    \includegraphics[trim={0cm 0cm 0cm 1.15cm}, clip, width=0.24\textwidth]{  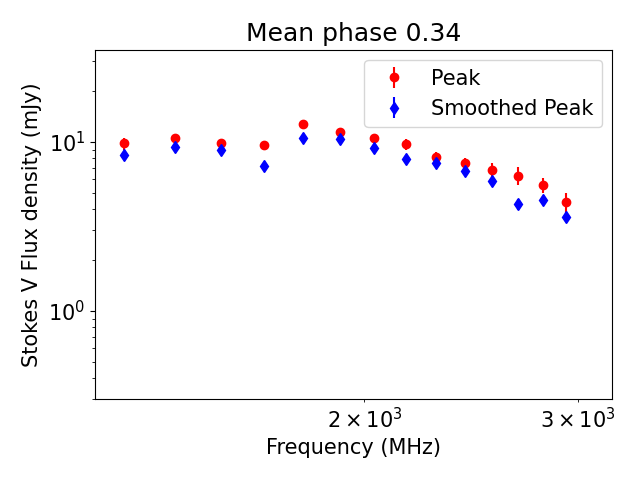}
    \includegraphics[trim={0cm 0cm 0cm 1.15cm}, clip, width=0.24\textwidth]{  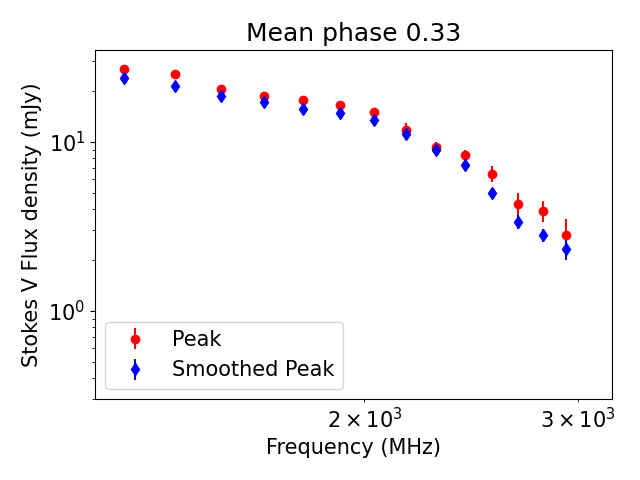}
    \includegraphics[trim={0cm 0cm 0cm 1.15cm}, clip, width=0.24\textwidth]{  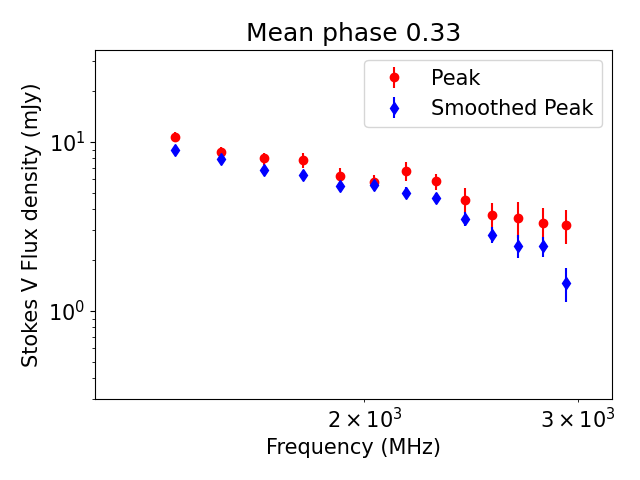}
    \includegraphics[trim={0cm 0cm 0cm 1.15cm}, clip, width=0.24\textwidth]{  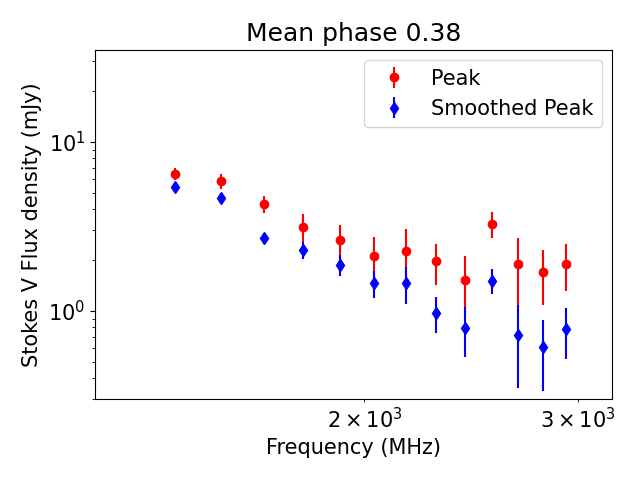}
    \caption{Peak Stokes V spectra (128 MHz spectral averaging and 2 min time averaging) for the leading pulse obtained with the two methods. \label{fig:spec_leading}}
\end{figure*}

\begin{figure*}
    \centering 
    \includegraphics[trim={0cm 0cm 0cm 1.15cm}, clip, width=0.24\textwidth]{  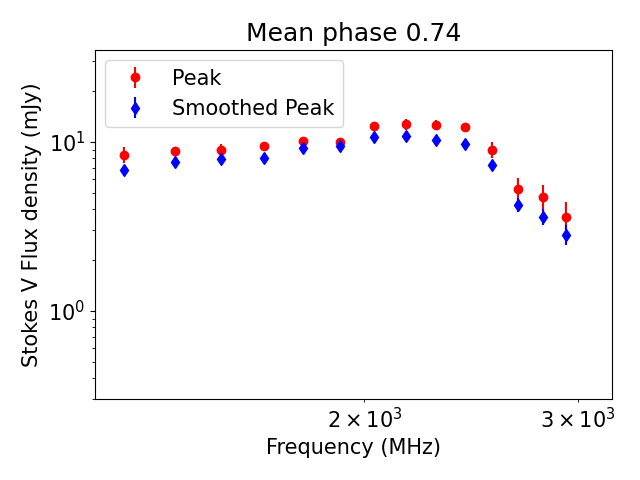}
    \includegraphics[trim={0cm 0cm 0cm 1.15cm}, clip, width=0.24\textwidth]{  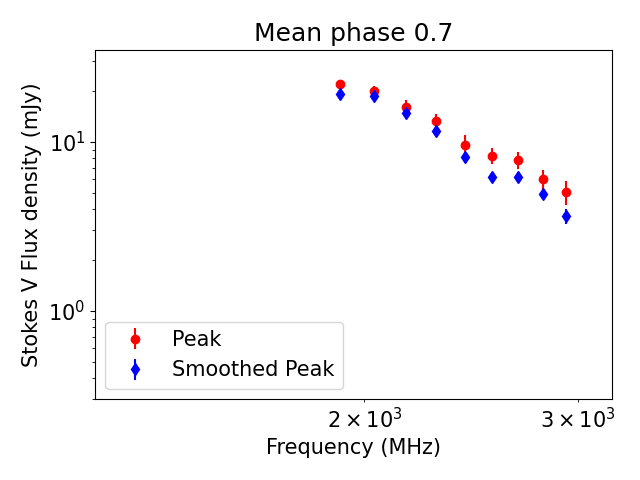}
    \includegraphics[trim={0cm 0cm 0cm 1.15cm}, clip, width=0.24\textwidth]{  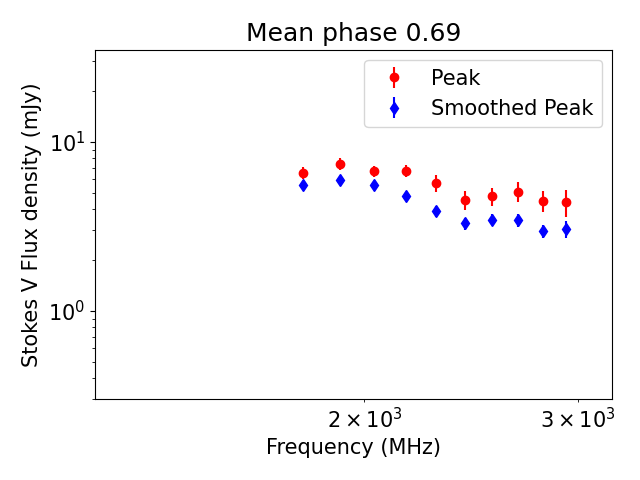}
    \includegraphics[trim={0cm 0cm 0cm 1.15cm}, clip, width=0.24\textwidth]{  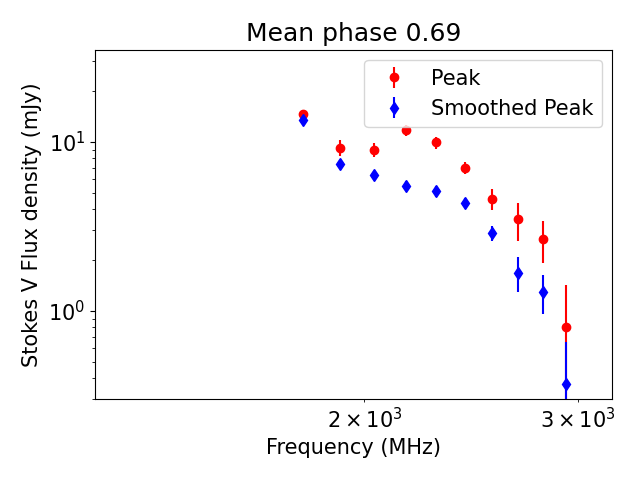}
    \includegraphics[trim={0cm 0cm 0cm 1.15cm}, clip, width=0.24\textwidth]{  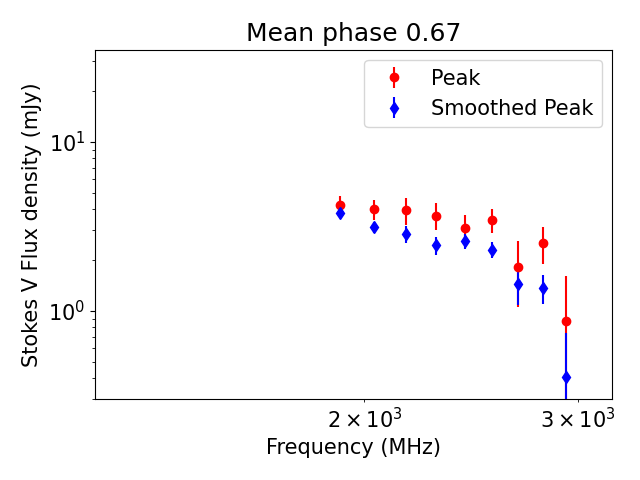}
    \includegraphics[trim={0cm 0cm 0cm 1.15cm}, clip, width=0.24\textwidth]{  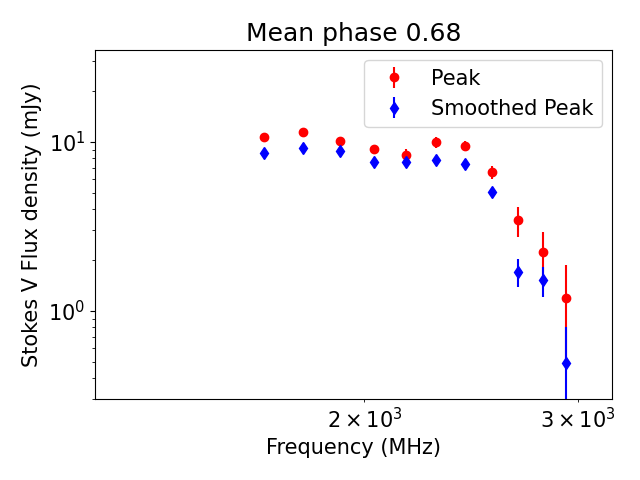}
    \includegraphics[trim={0cm 0cm 0cm 1.15cm}, clip, width=0.24\textwidth]{  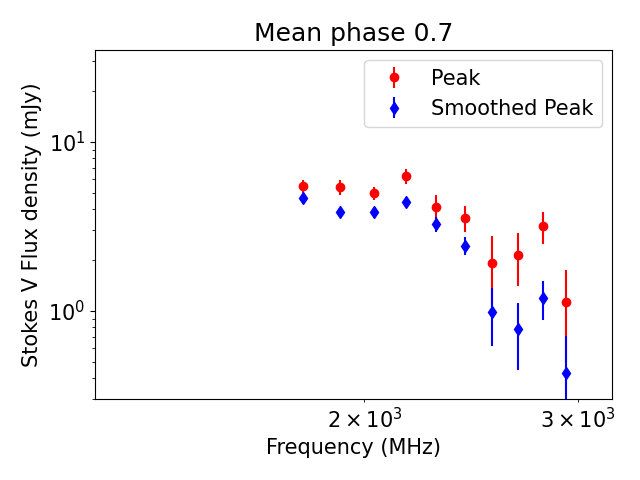}
    \includegraphics[trim={0cm 0cm 0cm 1.15cm}, clip, width=0.24\textwidth]{  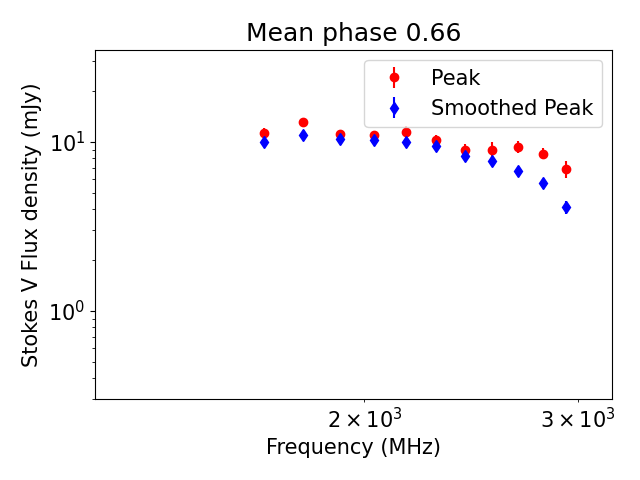}
    \includegraphics[trim={0cm 0cm 0cm 1.15cm}, clip, width=0.24\textwidth]{  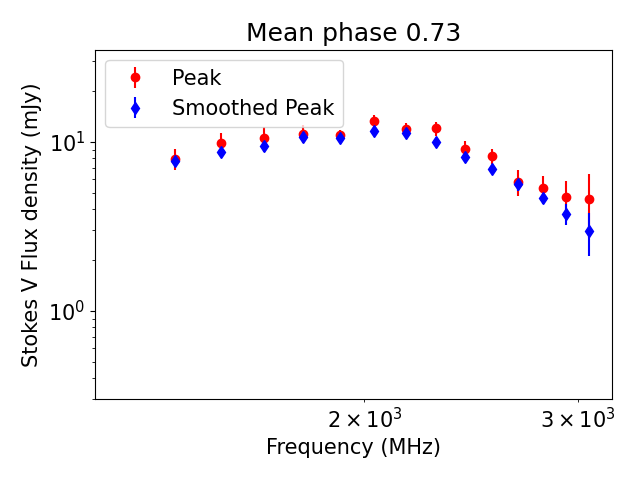}
    \includegraphics[trim={0cm 0cm 0cm 1.15cm}, clip, width=0.24\textwidth]{  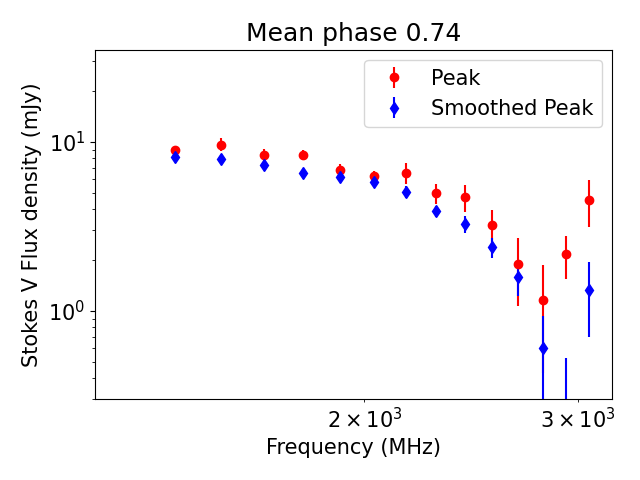}
    \includegraphics[trim={0cm 0cm 0cm 1.15cm}, clip, width=0.24\textwidth]{  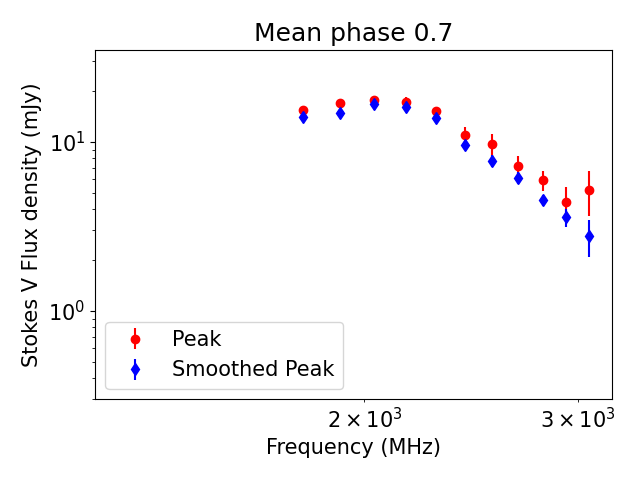}
    \includegraphics[trim={0cm 0cm 0cm 1.15cm}, clip, width=0.24\textwidth]{  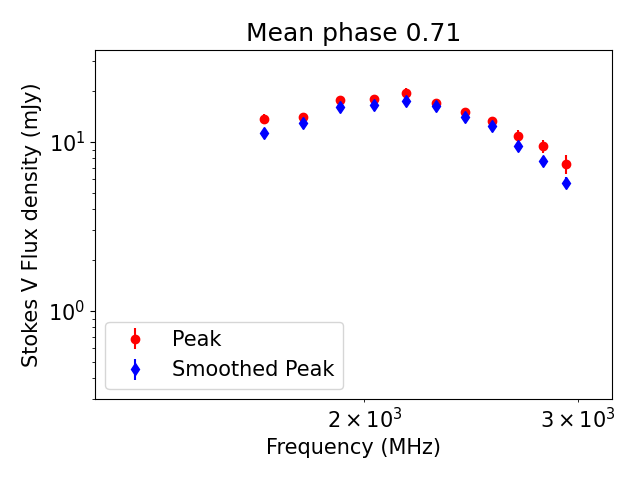}
    \includegraphics[trim={0cm 0cm 0cm 1.15cm}, clip, width=0.24\textwidth]{  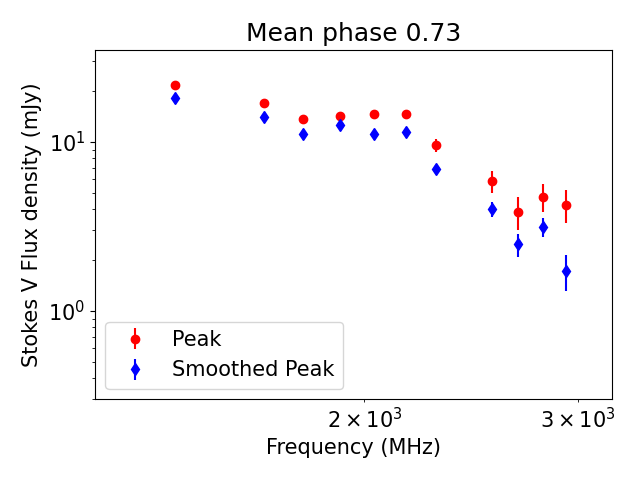}
    \includegraphics[trim={0cm 0cm 0cm 1.15cm}, clip, width=0.24\textwidth]{  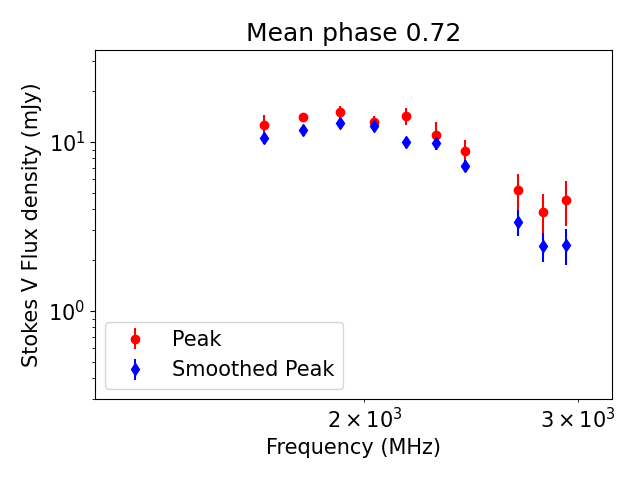}
    \includegraphics[trim={0cm 0cm 0cm 1.15cm}, clip, width=0.24\textwidth]{  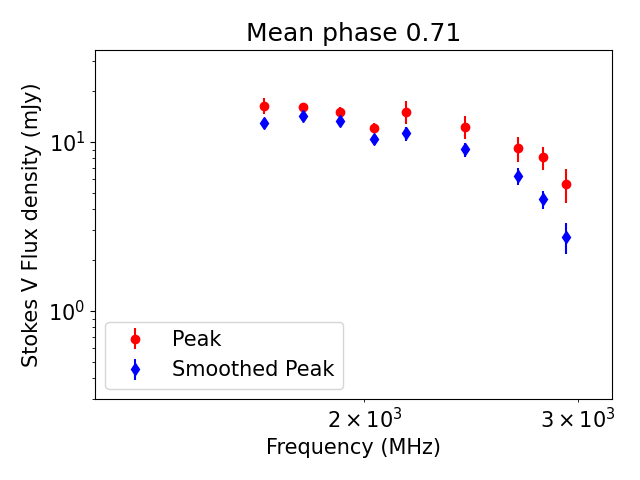}
    \includegraphics[trim={0cm 0cm 0cm 1.15cm}, clip, width=0.24\textwidth]{  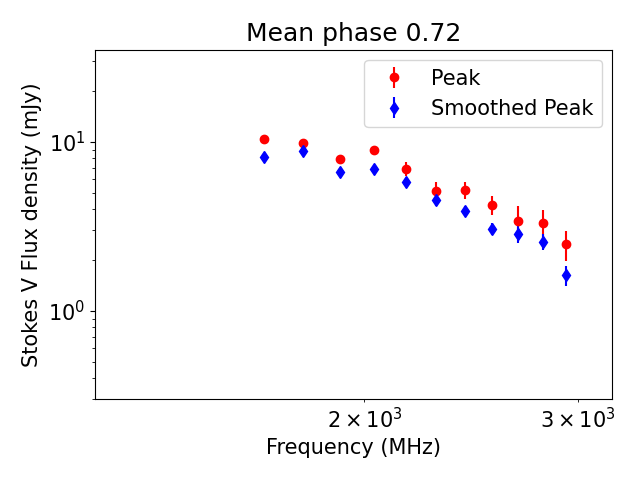}
    \includegraphics[trim={0cm 0cm 0cm 1.15cm}, clip, width=0.24\textwidth]{  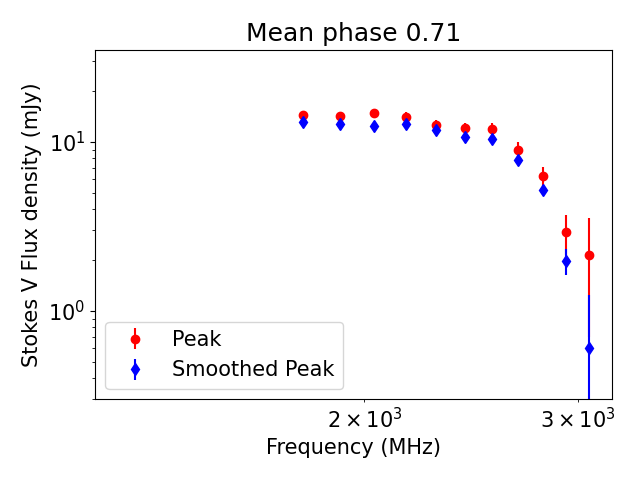}
    \includegraphics[trim={0cm 0cm 0cm 1.15cm}, clip, width=0.24\textwidth]{  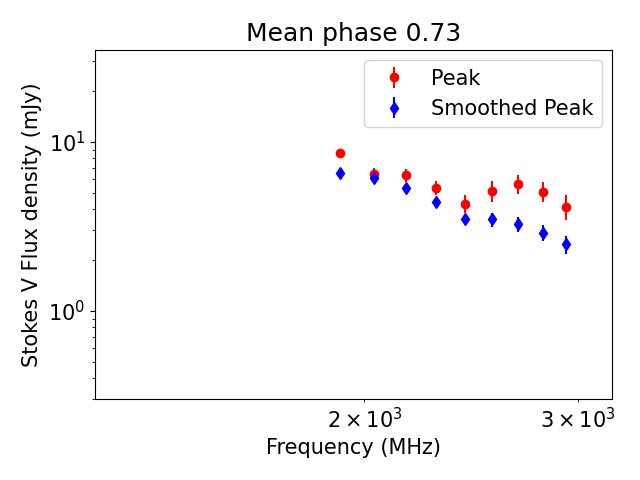}
    \includegraphics[trim={0cm 0cm 0cm 1.15cm}, clip, width=0.24\textwidth]{  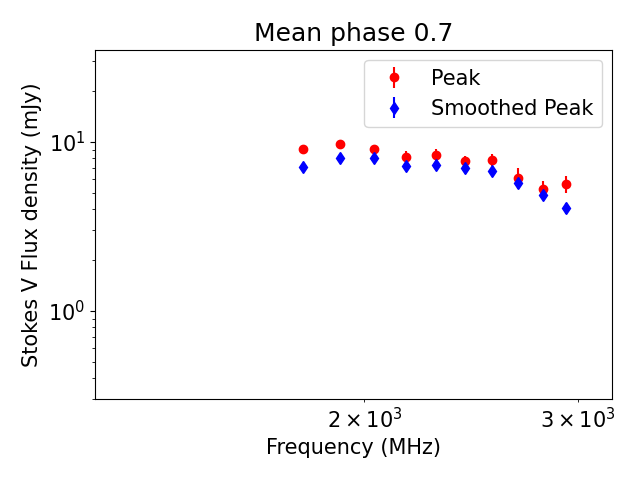}
    \caption{Peak Stokes V spectra (128 MHz spectral averaging and 2 min time averaging) for the trailing pulse. \label{fig:spec_trailing}}
\end{figure*}

\bibliography{das}{}
\bibliographystyle{aasjournalv7}



\end{document}